\documentclass[showpacs,preprintnumbers,amsmath,
amssymb,nofootinbib]{revtex4}
\usepackage{graphicx,bm}
\usepackage{dcolumn}    
 \usepackage{epsf}
\input epsf
\newcommand{\beq}{\begin{equation}}
\newcommand{\beqa}{\begin{eqnarray}}
		  \newcommand{\eeq}{\end{equation}}
\newcommand{\eeqa}{\end{eqnarray}}

\newcommand{\lsim}{\lesssim}
\newcommand{\gsim}{\gtrsim}

\newcommand{\vect}[1]{\mbox{\boldmath${#1}$}}
\newcommand{\lmk}{\left(}
\newcommand{\rmk}{\right)}
\newcommand{\lnk}{\left\{ }
\newcommand{\rnk}{\right\} }
\newcommand{\lkk}{\left[}
\newcommand{\rkk}{\right]}
\newcommand{\lla}{\left\langle}
\newcommand{\p}{\partial}
\newcommand{\rra}{\right\rangle}

\newcommand{\vex}{{\vect x}}

\newcommand{\ven}{\vect n}
\newcommand{\vem}{\vect m}

\newcommand{\veu}{{\vect u}}
\newcommand{\vev}{{\vect v}}

\newcommand{\ved}{{\vect d}}
\newcommand{\ve}{{\vect e}}

\begin{document}

\title{Polarization analysis of gravitational-wave backgrounds 
from the correlation signals of ground-based interferometers: 
measuring a circular-polarization mode } 
\author{Naoki Seto$^1$ and Atsushi Taruya$^2$}
\affiliation{$^1$Division of Theoretical Astronomy, National
Astronomical Observatory of Japan, 2-21-1 Osawa, Mitaka, Tokyo, Japan
181-8588, Japan\\
$^2$Research Center for the Early Universe, School of Science, 
The University of Tokyo, Tokyo 113-0033, Japan
}
\date{\today}
\begin{abstract}
 The Stokes $V$ parameter characterizes asymmetry of amplitudes between 
 right- and left-handed waves, and non-vanishing value of the $V$ parameter 
 yields a circularly polarized signal. Cosmologically, $V$ parameter may be 
 a direct probe for parity violation in the  universe. In this paper, 
 we theoretically investigate a measurement of this parameter, particularly 
 focusing on the gravitational-wave backgrounds observed via ground-based 
 interferometers. 
 In contrast to the traditional analysis that only considers the total 
 amplitude (or equivalently $\Omega_{\rm GW}$), the signal analysis including 
 a circular-polarized mode has a rich structure due to the 
 multi-dimensionality of target parameters. We show that, by using the 
 network of next-generation detectors, separation between polarized and 
 unpolarized modes can be performed with small statistical loss induced 
 by their correlation. 
\end{abstract}

\maketitle

\section{Introduction}

Because of the extremely weak signal, a direct detection of gravitational 
waves is a technically challenging issue, and  
we have not yet succeeded the direct detection despite extensive
efforts.  Nevertheless, 
the weakness of the gravitational interaction may be a great 
advantage for astronomy and cosmology, because gravitational waves 
can propagate to us from very early universe almost without scattering 
and absorption \cite{Thorne_K:1987,Cutler:2002me}. In this respect, 
stochastic background of gravitational waves is 
one of the most important targets for gravitational wave astronomy 
\cite{Allen:1996vm}. If detected, the stochastic background will serve 
as an invaluable fossil to study the physics at extremely high-energy 
scale for which other methods cannot be attainable.

Over the last decade, sensitivity of gravitational wave detectors 
to the stochastic background has drastically improved. We will soon 
reach at the sensitivity level $\Omega_{\rm GW}\lsim 
10^{-5}$ around 100Hz \cite{Abbott:2006zx}, where $\Omega_{\rm GW}$ is 
the energy density of the gravitational waves normalized by critical 
density of the universe. This level is below the 
indirect cosmological constraints, such as derived from the observed 
abundance of light elements \cite{Allen:1996vm} (see \cite{Smith:2006nka} 
for the constraints from cosmic microwave background), and 
in this sense, gravitational wave detectors will provide a unique 
opportunity to directly constrain the early universe. 
In order to further get a stringent constraint and/or valuable 
information from the next-generation detectors, one important approach 
is to improve the statistical analysis of gravitational wave backgrounds. 
So far,  most of theoretical studies on the gravitational-wave backgrounds 
have been directed  to its energy  spectrum (for its anisotropies, see,  
{\it e.g.,} \cite{Giampieri:1997ie}). The authors recently provided a 
brief sketch for measurement of the Stokes $V$ parameter of the 
gravitational-wave background via correlation analysis of 
ground-based detectors \cite{Seto:2007tn} (see \cite{Lue:1998mq} for cosmic 
microwave background and \cite{Seto:2006hf}
for space gravitational wave detectors such as  LISA \cite{lisa} or 
BBO/DECIGO \cite{bbo,Seto:2001qf}).  The Stokes $V$ parameter may be 
basic observable to quantify the parity violation process. One of such 
parity violation process is through the Chern-Simons term that might be 
originated from string theory \cite{Alexander:2004us}. This paper is 
a follow-up study to the preceding short report. In 
addition to detailed explanations and supplementary materials to the
previous paper, we developed a new statistical framework to deal with
multiple parameters of gravitational wave backgrounds, 
and we specifically 
applied it to simultaneous estimation of amplitudes of both the
unpolarized and polarized modes of the gravitational-wave background.

This paper is organized as follows. In section \ref{sec:circular_pol}, 
we describe polarization decomposition of 
a gravitational-wave  background,  and define its Stokes $V$ 
parameter. The basic framework to treat polarized gravitational waves 
is essentially the same one as in the case of electromagnetic waves 
\cite{radipro}. 
In section \ref{sec:overlap_func}, we explain the correlation analysis 
for the gravitational-wave background and
introduce the overlap functions that characterize sensitivities to the
polarized and unpolarized modes.  Then, we discuss basic properties of
the overlap functions, and calculate them for the planed 
next-generation detectors, such as advanced LIGO. In section 
\ref{sec:broadband_SNR}, broadband analysis of the gravitational-wave 
background is studied, taking into account the measurement noises for 
each detector.
In section \ref{sec:separation}, we discuss how well we can separately 
measure the polarized and unpolarized modes.  In contrast to the traditional 
arguments only for the unpolarized mode, the situation considered here 
is more complicated. We provide a statistical framework to analyze 
multiple parameters of the stochastic background with correlation analysis. 
Finally, section \ref{sec:summary} is a brief summary of this paper. 
Appendix \ref{sec:tensor_analysis} presents the derivation 
for the expressions of the overlap functions.  This geometrical 
derivation is similar to that given in Ref.\cite{Flanagan:1993ix}. 
In appendix \ref{sec:PDF_for_corr}, we discuss the probability distribution 
functions (PDFs) of basic observational quantities with correlation 
analysis. In appendix \ref{sec:derivation_optimal_SNR}, we derive the 
formal expressions for optimal signal-to-noise ratios for detectors more 
than four. In appendix \ref{sec:moon}, we comment on the surface of the Moon as potential sites for laser interferometers.

\section{Circular polarization}
\label{sec:circular_pol}

Let us first describe the polarization states of stochastic gravitational
waves. We consider a plane wave expansion of gravitational-wave 
backgrounds as 
\beq
h_{ij}(t,\vex)=\sum_{P=+,\times} \int^{\infty}_{-\infty} df \int_{S^2} d\ven~
h_P(f,\ven) e^{2\pi i f (-t+\ven \cdot \vex) } \ve^P_{ij}(\ven).\label{plane}
\eeq
Here, the bases for transverse-traceless tensor $\ve^P$ $(P=+,\times)$
are given as 
\beq
\ve^+_{}={\hat \ve}_\theta \otimes {\hat \ve}_\theta- {\hat \ve}_\phi
\otimes  {\hat \ve}_\phi,
\quad
\ve^\times_{}={\hat \ve}_\theta \otimes
{\hat 
\ve}_\phi+{\hat  
\ve}_\phi \otimes {\hat \ve}_\theta
\eeq
with the unit vectors $ {\hat \ve}_\theta$ and $ {\hat \ve}_\phi$ 
being normal to the propagation 
direction $\ven$ that are associated with a right-handed Cartesian 
coordinate: 
\beq
{\hat \ve}_\theta=(\cos\theta\cos\phi,\cos\theta\sin\phi,-\sin\theta),~~
{\hat \ve}_\phi=(-\sin\phi,\cos\phi,0).
\eeq
On the other hand, the random amplitude $h_P$ represents the mode 
coefficients and the statistical properties of it are characterized by the 
power spectral density given by
$\lla h_{P1}(\ven) h_{P2}^* (\ven') \rra$ $(P1,P2=+,\times)$
for two polarization modes as
\beq
\left( 
           \begin{array}{@{\,}cc@{\,}}
\lla h_{+}(f,\ven) h_{+}^* (f',\ven') \rra & 
\lla h_{+}(f,\ven) h_{\times}^* (f',\ven') \rra  \\
            \lla h_{\times}(f,\ven) h_{+}^* (f',\ven') \rra & 
\lla h_\times(f,\ven) h_{\times}^* (f',\ven') \rra  \\ 
           \end{array} \right)=\frac12
{\delta_{\rm D}^2(\ven-\ven')\delta_{\rm D}(f-f')}\left( 
           \begin{array}{@{\,}cc@{\,}}
           I(f,\ven)+Q(f,\ven) & U(f,\ven)-iV(f,\ven)  \\
            U(f,\ven)+iV(f,\ven) & I(f,\ven)-Q(f,\ven)  \\ 
           \end{array} \right), \label{matrix}
\eeq
with  delta functions $\delta_{\rm D}(\cdot)$ and the
notation $\lla \cdots \rra$ for an ensemble average. 
Here, the quantities $I,Q,U$ and $V$ are the Stokes parameters and are
real functions  of  direction $\ven$. 
Alternative to the linear polarization
bases $(\ve^+, \ve^\times)$, we may use the circular polarization bases
$(\ve^R, \ve^L)$ (right- and left-handed modes)
\beq
\ve^R=\frac{(\ve^++i\ve^\times)}{\sqrt2}, 
\quad
\ve^L=\frac{(\ve^+-i\ve^\times)}{\sqrt2 }
\eeq
 for the plane wave expansion (\ref{plane}). Two coefficients $h_{R,L}$ 
for the corresponding  modes  are given as 
\beq
h_R=\frac{(h_+-ih_\times)}{\sqrt2},~~~h_L=\frac{(h_++ih_\times)}{\sqrt2}.
\eeq
Then the covariance matrix is recast as
\beq
\left( \begin{array}{@{\,}cc@{\,}}
             \lla h_R(f,\ven) h_R(f',\ven')^* \rra  &
              \lla h_L(f,\ven) h_R(f',\ven')^* \rra \\
              \lla h_R(f,\ven) h_L(f',\ven')^* \rra &
              \lla h_L(f,\ven) h_L(f',\ven')^* \rra \\ 
           \end{array} \right)
=\frac12{\delta_{\rm D}({\ven-\ven'})^2\delta_{\rm D}({f-f'})}\left( 
           \begin{array}{@{\,}cc@{\,}}
           I(f,\ven)+V(f,\ven)  &
           Q(f,\ven)-iU(f,\ven)  \\ 
           Q(f,\ven)+iU(f,\ven)  &
           I(f,\ven)-V(f,\ven)  \\ 
           \end{array} \right).  \label{matrix2}
\eeq

With this expression, it is apparent that the real parameter $V$
characterizes the 
asymmetry of amplitudes between right- and  left-handed waves, 
while the parameter $I(\ge |V|)$ represents their total amplitude.
For example, if we can observationally establish $V>0$, the background 
is dominated by right-handed waves. Since the parity transformation 
interchanges the two polarization modes, the asymmetry is closely related 
to parity violation process (see {\it e.g.} 
\cite{Alexander:2004us,Kahniashvili:2005qi} for recent theoretical studies).  
Therefore, we may detect signature of parity violation in the early
universe by analyzing the $V$ parameter of gravitational-wave 
backgrounds.  This is the basic motivation of this paper.

Since the two parameters $I$ and $V$ have spin 0, 
their angular dependence can be expanded by the standard (scalar) spherical
harmonics $Y_{\ell m}$:
\beq
I(f,\ven)=\sum_{\ell=0}^\infty\sum_{m=-\ell}^\ell I_{\ell m}(f)Y_{\ell m}(\ven),
\quad
V(f,\ven)=\sum_{\ell=0}^\infty\sum_{m=-\ell}^\ell V_{\ell m}(f)Y_{\ell m}(\ven).
\eeq
On the other hand, the combinations $Q\pm iU$ describe the
linear polarization and have spin $\pm 4$ reflecting spin-2 nature of
gravitational waves. They are expanded with the
spin-weighted spherical harmonics as
\beq
(Q+i\,U)(f,\ven)=\sum_{\ell=4}^\infty\sum_{m=-\ell}^\ell P^+_{\ell m}(f) 
{}_4Y_{\ell m}(\ven),
\quad
(Q-i\,U)(f,\ven)=\sum_{\ell=4}^\infty\sum_{m=-\ell}^\ell P^-_{\ell m}(f) 
{}_{-4}Y_{\ell m}(\ven).
\eeq
Note that $Q\pm i\,U$ do not have monopole components ($\ell=0$), because 
the linear modes introduce specific spatial directions.
Since the observed universe is highly homogeneous and isotropic on large
spatial scales, it is reasonable to set the monopole modes of a 
cosmological stochastic background as our primary targets. 
Therefore, in this paper, we do not study the linear polarization $Q\pm
i\,U$. From  the same reason, we also 
neglect the directional dependence of the $I$ and $V$ modes.

Next, we describe the frequency dependence of the gravitational-wave 
background. To characterize the gravitational waves in the 
cosmological context, rather than the spectral density, 
the normalized logarithmic energy density of the stochastic background, 
$\Omega_{\rm GW}(f)$, is frequently used in the literature 
\cite{Flanagan:1993ix,Allen:1997ad}. 
The density $\Omega_{\rm GW}(f)$ is defined by the spectral density $I$ as
\beq
\Omega_{\rm GW}(f)=\frac{4\pi^2 f^3}{\rho_c} I(f),
\eeq 
where $\rho_{\rm c}(=3H_0^2/8\pi$, $H_0=70h_{70}$km/sec/Mpc: 
the Hubble parameter) is the critical density of the universe. 
We also define the polarization degree by $\Pi(f)=V(f)/I(f)$. 
In terms of the quantities $\Omega_{\rm GW}(f)$ and $\Pi(f)$, 
the asymmetry parameter $V$ is expressed as 
\beq
V(f)=\frac{\rho_{\rm c}}{4\pi^2 f^3} \Omega_{\rm GW}(f) \Pi(f).
\eeq

\section{Overlap functions for ground-based detectors}
\label{sec:overlap_func}

This section discusses the overlap functions for 
correlation signals as the basic ingredient for correlation analysis of 
gravitational-wave background. In Sec.\ref{subsec:formulation}, 
the definition and the analytic formula for overlap functions are 
given. Subsequently, Sec.\ref{subsec:special}, \ref{subsec:same_plane} 
and \ref{subsec:optimal_config} discuss special or limiting cases 
for overlap functions in order to understand their geometrical properties. 
After describing some mathematical properties in 
Sec.\ref{subsec:functions_Theta}, we evaluate the overlaps functions for 
specific pairs of five detectors in Sec.\ref{subsec:overlap_specific}.

\subsection{Formulation}
\label{subsec:formulation}

Let us begin by considering how we can detect the monopole 
components of the $I$ and $V$ modes with laser interferometers. 
Response $H_a$ of a detector $a$ at $\vex_a$  is written as
\beq
H_a(f) =\int_{S^2} d\ven  
\sum_{P=+,\times} h_P(f,\ven) F^P_{a}(\ven, f) e^{2\pi i f \ven\cdot\vex_a}. 
\eeq
The function  $F_a^P$ is   the beam pattern function and it
represents the response of the detector to each linear  polarization
mode.   Here we used the conventional linear polarization bases.

To distinguish the background signals from detector noises and to obtain  a
large signal-to-noise ratio, the correlation analysis with multiple
detectors is essential \cite{m87,Christensen:1992wi,Flanagan:1993ix,
Allen:1997ad}. We define the correlation $C_{ab}(f)$ of data streams 
obtained from two detectors $a$ and $b$ as
\beq
 \lla  H_a(f) H_b(f')^* \rra \equiv C_{ab}(f) \delta_{\rm D}(f-f').
\eeq
Keeping the monopole contribution only,  its expectation 
value is written as
\beq
 C_{ab}(f)=\frac{8\pi}5\lkk 
\gamma_{I,ab}(f)I(f)+\gamma_{V,ab}(f)V(f) \rkk,
\eeq 
where $\gamma_I$ is the overlap  function and given by \cite{Flanagan:1993ix,
Allen:1997ad}
\beq
\gamma_{I,ab}(f)=\frac{5}{8\pi}\int_{S^2} d\ven  \lkk  \lnk
F_a^+F_{b}^{+*}+
F_a^\times F_{b}^{\times*} \rnk e^{iy \ven\vem} \rkk, \label{gi11}
\eeq
with rewriting $\vex_a-\vex_b=D \vem$ ($D$: distance, $\vem$: unit vector) 
and $y\equiv2\pi fD/c$.  The variable $y$ represents the phase difference at
two cites $a$ and $b$ for waves with  a propagation direction $\vem$.  
Similarly, the function $\gamma_{V,ab}(f)$ is given by
\beq
\gamma_{V,ab}(f)=\frac{5}{8\pi}\int_{S^2} d\ven  \lkk i \lnk
F_a^+F_{b}^{\times*}-
F_a^\times F_{b}^{+*} \rnk e^{iy \ven\vem}  \rkk. \label{gv11}
\eeq
Two functions $\gamma_I$ and $\gamma_V$ are purely determined by relative 
configuration of two detectors.

Here, we summarize the response of ground-based L-shaped interferometer $a$.  
We assume that two arms of the next-generation interferometer have equal 
length with opening angle of $90^\circ$.  
We denote the unit vectors for the directions of its two arms as $\veu$ and
$\vev$. At the frequency regime where the wavelength of the incident
gravitational wave is much longer than the armlength,  the beam pattern
function takes a  simple form as
\beq
F_a^P=\ved_a:\ve^P(\ven),  
\label{eq:response}
\eeq
where the colon represents a double contraction and  the detector  tensor
$\ved_a$ is given by 
\beq 
\label{detten}
\ved_a=\frac{({\veu}_a \otimes {\veu}_a- {\vev}_b
\otimes  {\vev}_b)}2.
\eeq
In reality,  there might be some 
exceptional cases that the opening angle of two arms is slightly different 
from $90^\circ$ such as GEO600, whose opening angle is $94.3^\circ$ 
\cite{Willke:2002bs}. However, the response of such a detector
can be treated as the one of the right-angled interferometer (see {\it eg.} \cite{lisa} for the case with LISA).

Note that from equations (\ref{eq:response}) and (\ref{detten}),  we have
\beq
\frac{5}{8\pi}\int_{S^2} d\ven  \lkk  
F_i^+F_{i}^{+*}  \rkk=\frac{5}{8\pi}\int_{S^2} d\ven  \lkk  
F_i^\times F_{i}^{\times*}  \rkk=\frac12,  
\eeq
and Schwartz inequality implies 
\beq
-1\le \gamma_{I,ab}\le 1,~~~-1\le \gamma_{V,ab}\le 1.
\eeq

In Table \ref{tab:location_of_detectors}, we list positions and orientations 
of the ongoing (and planned) kilometer-size interferometers, 
AIGO \cite{aigo}, LCGT \cite{Kuroda:1999vi}, LIGO-Hanford,
LIGO-Livingston \cite{Abramovici:1992ah}, and Virgo \cite{Acernese:2002bw}.
As a reference, we also list two sub-kilometer-size
interferometers, TAMA300 \cite{Ando:2001ej} and GEO600 \cite{Willke:2002bs}.  
We use a spherical coordinate system $(\theta,\phi)$ with 
which the north pole is at $\theta=0^\circ$, and  $\phi$
 represents longitude. The orientation $\alpha$ is the angle between 
the local east direction and the bisecting line of  two arms of each detector
measured  counterclockwise.
Since the beam pattern functions have spin-2 character with respect to the 
rotation of detector, meaningful information here is the angle $\alpha$ 
module $90^\circ$. In what follows, we mainly focus on the first 
five detectors in Table \ref{tab:location_of_detectors} with their 
abbreviations (A,C,H,L,V) and with $R=R_{\rm E}=6400$km for the radius of 
the Earth \footnote{We use the roman V for Virgo detector and the italic $V$ 
for the polarization mode.}, but in section \ref{subsec:antipodal}, 
we also discuss the detectors placed on the Moon 
as an exceptional but interesting case (see also appendix \ref{sec:moon}).

\begin{table}[!bth]
\begin{tabular}{lccc}
\hline\hline
 ~ & $\theta$ & $\phi$ & $\alpha$ \\
\hline
\ AIGO ({\bf A}) & $121.4$ & $115.7$ & $-45.0$\\
\ LCGT ({\bf C}) & $53.6$ & $137.3$ & $70.0$ \\
\ LIGO\ Hanford ({\bf H}) & $43.5$ & $-119.4$ & $171.8$ \\
\ LIGO\ Livingston ({\bf L}) & $59.4$ & $-90.8$ & $243.0$ \\
\ Virgo ({\bf V}) & $46.4$ & $10.5$ & $116.5$ \\
\hline
\ TAMA300 & $54.3$ & $139.5$ & $225.0$ \\
\ GEO600  & $47.7$ & $9.8$ & $68.8$ \\
\hline\hline
\end{tabular}
\caption{Positions $(\theta,\phi)$ and orientation angles $\alpha$ of
 detectors (in units of degree) on the Earth.   }
\label{tab:location_of_detectors}
\end{table}

\begin{figure}
  \begin{center}
\epsfxsize=5cm
\epsffile{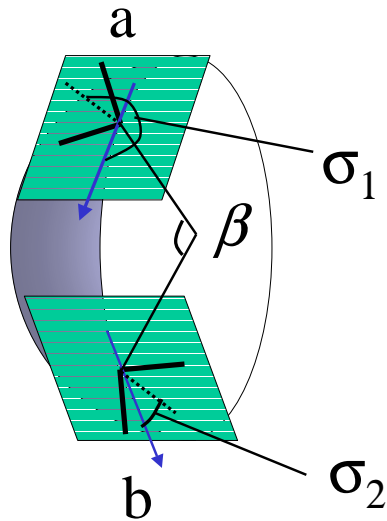}
 \end{center}

\vspace*{-0.5cm}

  \caption{  Detector planes are tangential to a sphere. Two detectors
 $a$ and $b$ are separated by the angle $\beta$ measured from the center
 of the sphere. The angles
 $\sigma_1$ and $\sigma_2$ describe the orientation of bisectors of
 interferometers in a counter-clockwise manner relative to the great 
 circle joining two sites. 
 }
\label{fig:f1}
\end{figure}

For the monopole modes of the stochastic background, only the
relative configuration of two detectors is relevant with the correlation
$C_{ab}$ and we do not need to deal with their overall rotation. 
Therefore, without loss of generality, their
configuration is characterized by the three angular parameters
$(\beta,\sigma_1,\sigma_2)$, shown in Figure \ref{fig:f1} 
\cite{Flanagan:1993ix}.   Here, $\beta$ is the separation angle between 
two detectors measured from the 
center of the Earth. The angle $\sigma_1$ ($\sigma_2$) is the
orientation of the bisector of two arms of the detector $a$ ($b$
respectively) measured in counter-clockwise manner relative to the 
great circle connecting $a$ and $b$. Their distance is given by
$D=2R_{\rm E} \sin(\beta/2)$ that determines a characteristic frequency 
$f_{\rm D}\equiv c/D$ for the overlap functions. 
Following \cite{Flanagan:1993ix}, we define the angles 
\beq
\Delta\equiv \frac{({\sigma_1+\sigma_2})}2,
\quad
\delta\equiv \frac{({\sigma_1-\sigma_2})}2,
\eeq
and  the geometrical information for possible pairs made from the five 
detectors are summarized in Table \ref{tab:table2}.

\begin{table}[!tbh]
\begin{ruledtabular}
\begin{tabular}{l|c|c|c|c|c}
  & A &  C & H & L & V \\
\hline
\ AIGO (A) & $*$ & $70.8,\, 58.1,\,31.4$ & $135.6,\,53.7,\,45.1$  
&$157.3,\,38.0,\,2.08$   &$121.4,\,19.2,\,60.8$ \\
\hline
\ LCGT (C) & $-0.61,\,-0.58,\,0.81$ & $*$  & $72.4,\,89.1,\,25.6$  
& $99.2,\,42.4,\,68.1$  &  $86.6,\,28.9,\,5.6$ \\
\hline
\ LIGO\ Hanford (H) & $-0.82,\,-1.00,\,-0.007$  & $1.0,\,-0.21,\,0.98$  
& $*$ &  $27.2,\,45.3,\,62.2$ & $79.6,\,61.8,\, 55.1$ \\
\hline
\ LIGO\ Livingston (L) &$-0.88,\,0.99,\,0.15$  &$-0.98,\,0.04,\,-1.0$  
& $-1.00,\,-0.36,\,-0.93$   & $*$ & $76.8,\,26.7,\,83.1$ \\
\hline
\ Virgo (V) & $0.23,\,-0.45,\,-0.89$  & $-0.43,\,0.92,\,0.38$ 
& $-0.43,\,-0.76,\,-0.65$  & $-0.29,\,0.89,\,-0.46$ &  $*$
\end{tabular}
\end{ruledtabular}
\caption{Upper right: angle parameters 
$(\beta,\delta,\Delta)$ for each pair of detectors in units of degree. 
Lower left: numerical values $(\cos(4\delta), \cos(4\Delta),\sin(4\Delta))$ 
for each pair of detectors.} 
\label{tab:table2}
\end{table}

The angular integral (\ref{gi11}) can be performed analytically 
 with explicit forms of the pattern functions, and  we get
\beq
\gamma_{I,ab}= \Theta_1(y,\beta)\cos(4\delta)+
\Theta_2(y,\beta) \cos(4\Delta), \label{gi}
\eeq
with
\beq
 \Theta_1(y,\beta)=\cos^4\lmk\frac{\beta}2  \rmk \lmk j_0+\frac57
j_2+\frac{3}{112} j_4 \rmk ,
\eeq
and 
\beq
 \Theta_2(y,\beta)=\lmk -\frac38 j_0+\frac{45}{56}
j_2-\frac{169}{896} j_4 \rmk
+\lmk \frac12 j_0-\frac57j_2-\frac{27}{224}j_4 \rmk \cos\beta
+ \lmk-\frac18 j_0-\frac5{56}j_2-\frac{3}{896}j_4  \rmk \cos(2\beta).
\eeq
The function $j_n$ is the $n$-th spherical Bessel function with its argument
\beq
y\equiv\frac{2\pi f D}{c}=\frac{4\pi f R_{\rm E}}{c}\,
\sin\left(\frac{\beta}{2}\right).
\eeq
The expression (\ref{gi}) coincides with the formula (4.1) in Ref.
\cite{Flanagan:1993ix}.

In  a similar manner, the overlap function for the $V$ mode is
given by 
\beq
\gamma_{V,ab}=\Theta_3(y,\beta)\sin(4\Delta) 
\label{gv}
\eeq
with
\beq
\Theta_3(y,\beta)=-\sin\lmk \frac{\beta}2 \rmk \lkk \lmk-j_1+\frac78
j_3  \rmk + \lmk j_1+\frac38 j_3 \rmk\cos\beta   \rkk.
\eeq
Note that the dependence of the angles $\delta$ and $\Delta$ on the 
overlap functions (\ref{gi}) and (\ref{gv}) can be 
deduced from the symmetric reasons \cite{Seto:2007tn}.

In appendix \ref{sec:tensor_analysis}, we present a brief sketch to derive 
the expressions $\gamma_I$ and $\gamma_V$, using the symmetries of 
tensorial structure. Since our primary interest here is the dependence 
on the frequency $f$ and the angle $\beta$, we mainly use the set of
the variables $(f,\beta)$, instead of $(y,\beta)$.

\subsection{Special cases and asymptotic profiles}
\label{subsec:special}

In order to get a physical insight into the overlap functions, 
it is instructive to consider geometrically simple 
configurations for two detectors. When a pair of detectors 
are placed on the same plane ($\beta=0^\circ$) and at the 
same position ($D=0$), 
we have the identity $(\Theta_1,\Theta_2)=(1,0)$ and  thus 
$\gamma_{I,ab}=\cos(4\delta)$. In contrast, for $V$ mode, 
we obtain $\gamma_{V,ab}=0$ for the coplanar configuration 
$(\beta=0^\circ)$ and this is even true 
with finite separation $D\ne0$. The reason for this is explained in next
subsection. Equation (\ref{gi}) and the identity 
$\Theta_2(y,0^\circ)=0$ indicates that 
the function $\gamma_I$ depends very weakly on the parameter $\Delta$ at 
small angle $\beta \ll 180^\circ$. For ground-based detectors, the 
functions $\Theta_i(y,\beta)$ depend on the angle $\beta$ also 
through the variable $y=4\pi R_{\rm E} f \sin(\beta/2)/c$. 
Taking into account this fact,  we obtain the following asymptotic
profiles at small $\beta$ (in unit of radian): 
\beq
\Theta_1=O(\beta^0),
\quad
\Theta_2=O(\beta^4),
\quad
\Theta_3=O(\beta^3).
\eeq

On the other hand, for pair of detectors located at 
antipodal positions ($\beta=\pi$), the overlap 
function $\gamma_{I,ab}$ does not depend on the parameter 
$\delta$ because of $\Theta_1(y,180^\circ)=0$. In this case, 
the asymptotic profiles become
\beq
\Theta_1=O((\pi-\beta)^4),
\quad
\Theta_2=O((\pi-\beta)^0),
\quad
\Theta_3=O((\pi-\beta)^0).
\eeq
Note that at $\beta=0$ and $\pi$, we have
\beq
\p_\beta \Theta_1=\p_\beta \Theta_2=\p_\beta \Theta_3=0.
\eeq

\subsection{Coplanar configuration}
\label{subsec:same_plane}

 An L-shaped detector measures difference of spatial deformation towards
 its orthogonal two arms.  This is purely geometrical measurement. If 
two detectors are placed on the same plane $z=0$, there is an apparent 
geometrical symmetry for the system with respect to the plane. Due to 
the mirror symmetry to the plane,  a right-handed wave 
coming from the direction 
$(n_x,n_y,n_z)$ and a left-handed wave from the direction
$(n_x,n_y,-n_z)$, provide an identical correlation signal, if they have the 
same frequency and amplitude. 
Therefore, for an isotropic background,   right-handed waves coming
from two directions  
$(n_x,n_y,\pm n_z)$ exactly cancel out  in the correlation signal. The 
same is true for left-handed waves.  As a result,  the symmetric
system has no sensitivity to the isotropic component of the $V$-mode 
\cite{Seto:2007tn,Seto:2006hf}.  
We can directly confirm this cancellation from the definition
(\ref{gv11}) and  the following  relations 
\beq
F^+_i(n_x,n_y,n_z)=F^+_i(n_x,n_y,-n_z),
\quad
F^\times_i(n_x,n_y,n_z)=-F^\times_i(n_x,n_y,-n_z), 
\eeq
which are easily derived from the symmetries of the polarization bases
$e^{+,\times}(\ven) $ \cite{Kudoh:2005as}.  
The cancellation of correlation signal is particularly important for 
setting orbits of space-based interferometers, 
such as BBO/DECIGO \cite{Seto:2006hf}. For detecting the 
monopole of the $V$-mode, it is essential to break the symmetric 
configuration.

\subsection{Optimal configuration}
\label{subsec:optimal_config}

In this subsection, we consider optimal configurations of two 
detectors $(a,b)$ for measuring the $I$ and $V$ modes of 
stochastic backgrounds. To investigate the optimized parameters for 
overlap functions, there are two relevant 
issues; maximization of the signals $\gamma_{I,ab}$ 
and $\gamma_{V, ab}$, and switching off either of them
($\gamma_{I,ab}=0$ or $\gamma_{I,ab}=0$) for their decomposition. 
For simplicity,  we consider how to set the second detector  $b$ relative to
the fixed first one $a$ for a given separation angle $\beta$. 
In this case, the sensitivities to the $I$- and $V$-modes 
are characterized by the remaining adjustable parameters, 
$\sigma_1$ and $\sigma_2$. 
The former determines the position of the detector $b$, while the latter 
specifies its orientation (see Fig.\ref{fig:f1}). 
Based on the expressions (\ref{gi}) and (\ref{gv}), 
one finds that there are three possibilities for the optimal detector 
orientation:  
\beq
\mathrm{Type\,\, I}: \quad \cos (4\Delta)=-\cos (4\delta)=\pm 1 
\eeq
or 
\beq
\mathrm{Type\,\, II}:\quad \cos (4\Delta)=\cos (4\delta)=\pm 1 
\eeq
to maximize the normalized SNR ${\it S}_{I,ab}$ \cite{Flanagan:1993ix}, 
and  
\beq
\mathrm{Type\,\,III}:\quad \cos{(4\Delta)}=\cos{(4\delta)}=0  
\eeq 
to erase the contribution from $I$-mode. 
The relative signs of the two functions $\Theta_1$ and $\Theta_2$
determine whether type I or type II is  the optimal choice.

\begin{figure}[tb]
\begin{center}
\epsfxsize=8cm
\epsffile{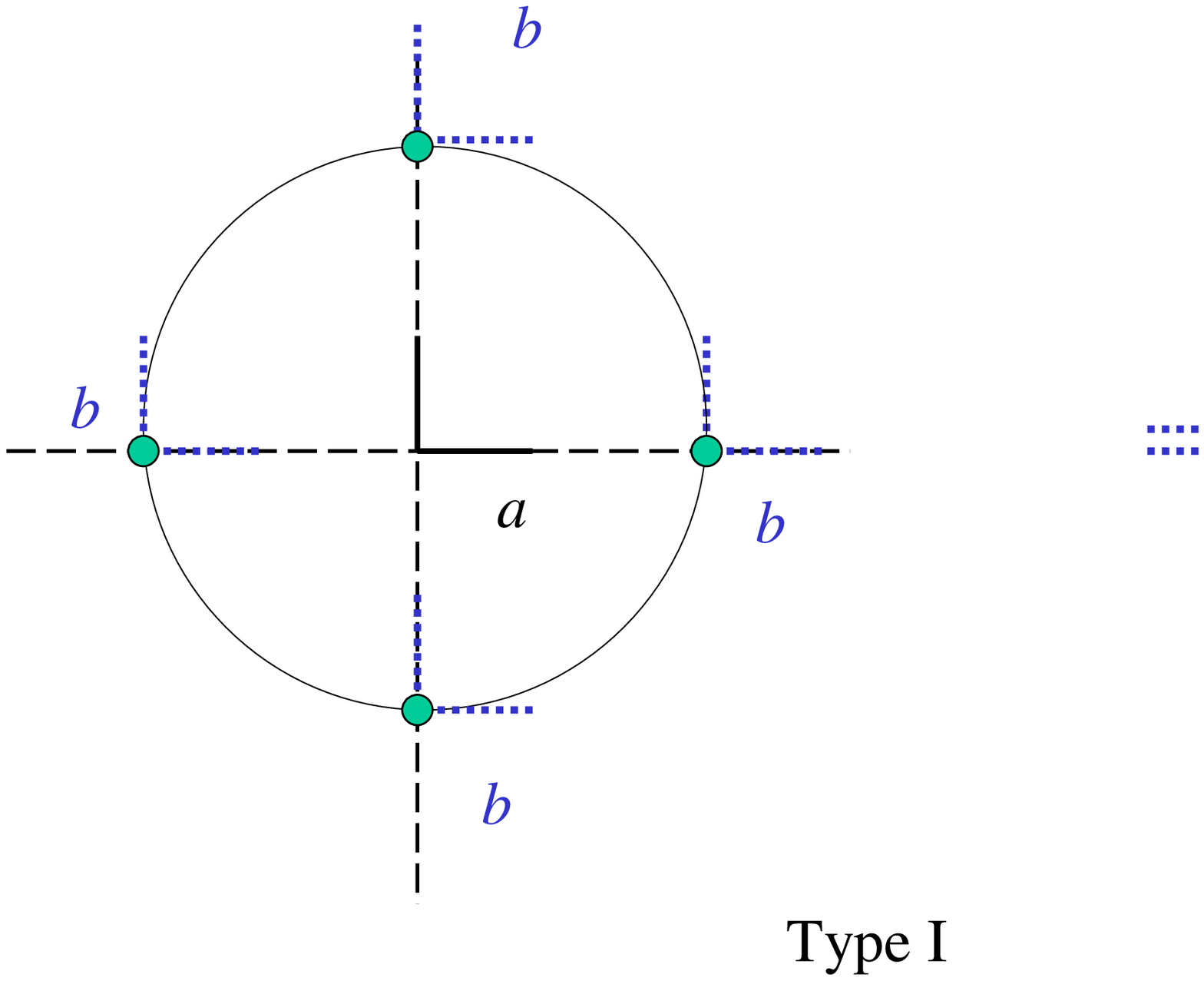}
\end{center}

\vspace*{-0.5cm}

\caption{Type I configuration with a given separation angle $\beta$. 
  Relative to a fixed L-shaped
 interferometer $a$, the second one must be placed on two great circles
 shown with long-dashed lines (left panel). We also have four
 equivalent detector orientations due to mod-$90^\circ$ freedom as
 shown in the right panel.} 
\label{fig:type1}

\vspace*{-1.0cm}

\begin{center}
\epsfxsize=12cm
\epsffile{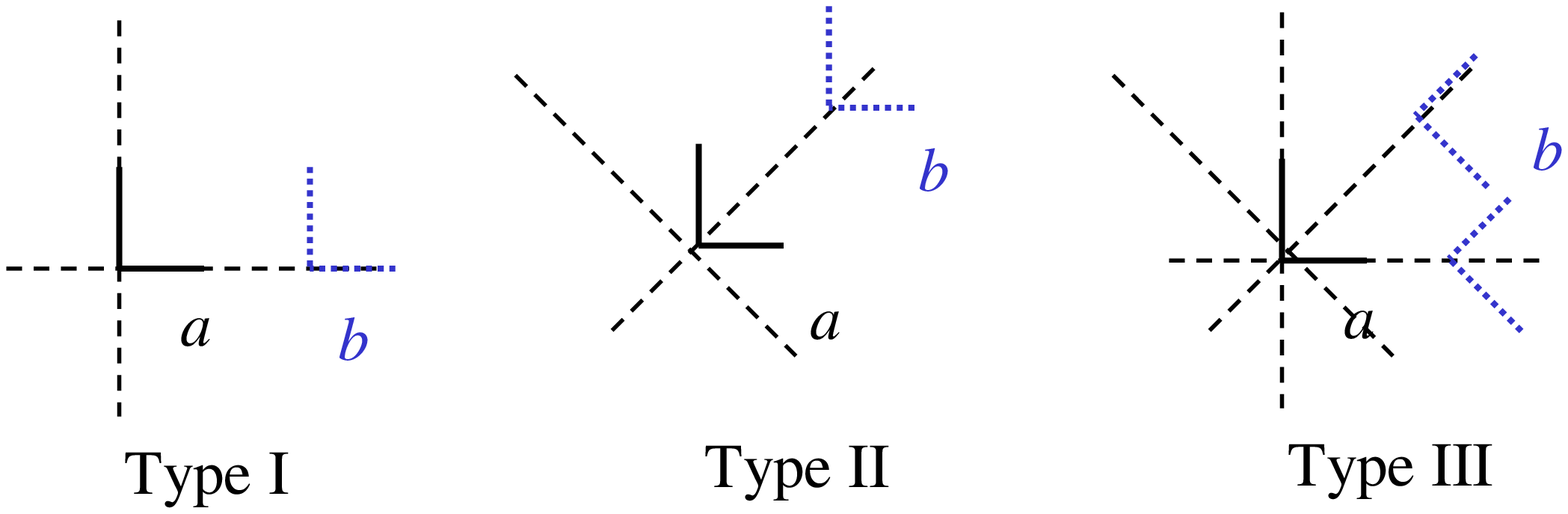}
\end{center}

\vspace*{-2cm}

\caption{ Position and orientation of the second detector $b$ relative to
 the fixed first one $a$. The long dashed lines are great circles
 passing the first one $a$.   } 
\label{fig:type}
\end{figure}
For type I, the solutions of the two angles $\sigma_{1,2}$ are 
$\sigma_1=\sigma_2=45^\circ$ 
(mod $90^\circ$) and the detector $b$ must be placed on one 
of the two great circles passing through the detector $a$, 
parallel to one of the two arms as shown in Figure \ref{fig:type1}.
For a given separation $\beta$,  there are four points for the  
cites of the detector $b$. At each point we have four equivalent 
orientations as shown in the right panel of Figure \ref{fig:type1}.
This is because response of an  L-shaped detector has mod-$90^\circ$
effective equivalence.  After all,  for a given  separation $\beta$, 
there are totally $4\times 4=16$ possible configurations of  detector 
$b$.

For type II, the second detector must reside in two great 
circles parallel or perpendicular to the bisecting line of 
each detector, as shown in Figure \ref{fig:type}. As in the case of type I,
with a given separation $\beta$ we have totally  16 candidates 
for  detector $b$.  At the  limits
$\beta \to 0^\circ$ and $\beta \to 180^\circ$, 
there are no essential differences   between types I and II.

 Similarly, the type III configuration is 
realized by placing the second detector on one of the 
four great circles defined for types I and II, with 
rotating $45^{\circ}$ relative to the first detector 
(see Fig.~\ref{fig:type}). 
In this case we have $8\times  4=32$ possible configurations
for  detector $b$. 
Note that the sensitivity to the $V$-mode is automatically switched off 
for the type I and II configurations and is conversely maximized for 
the type III configuration. This is because the function $\gamma_V$ is
proportional to $\sin (4\Delta)$.

\subsection{Basic properties of functions $\Theta_i$}
\label{subsec:functions_Theta}

In this subsection,  
specifically focusing on the detectors on the Earth with 
radius $R_{\rm E}$=6400km,  we study basic properties of the three
functions $\Theta_1$, $\Theta_2$ and $\Theta_3$ in some details. 
Note that in general, for a sphere  with radius $R_{\rm s}$, there is 
one characteristic frequency  $c/R_{\rm s}$ and our results for 
the Earth at frequency $f$ can be rescaled to those for the 
sphere with scaled frequency $(R_{\rm E}/R_{\rm s})\,f$ \footnote{  
This is easily deduced from the fact that the functions
$\Theta_1$,  $\Theta_2$ and $\Theta_3$ depend on  frequency $f$
only through the product $f\,R_{\rm E}$}. Hence, 
the result presented here may be interpreted as the one 
for an arbitrary sphere, including multiple detectors placed on the Moon. 
\begin{figure}[t]
\begin{center}
\includegraphics[width=6.5cm,clip]{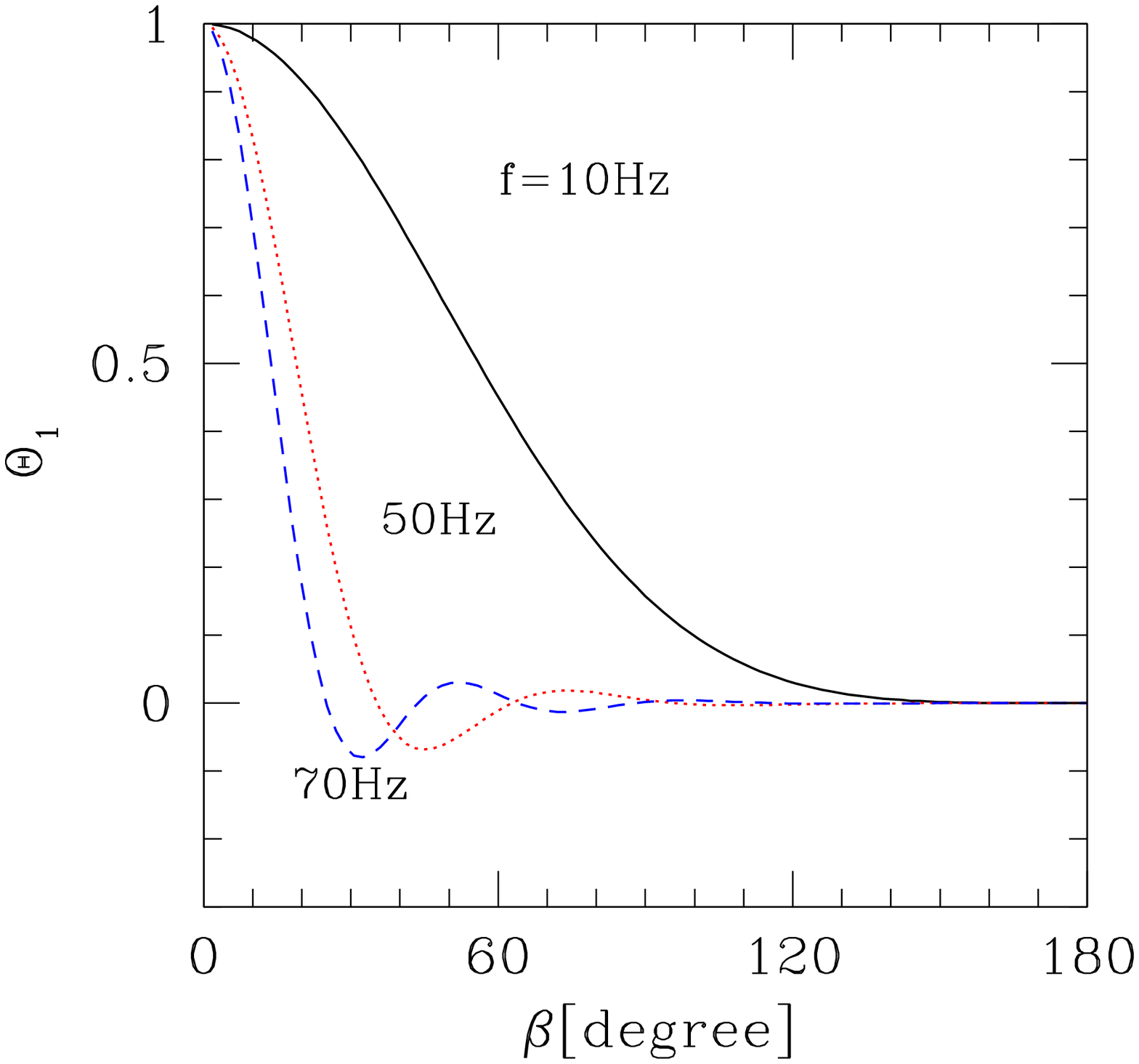}
\hspace*{0.5cm}
\includegraphics[width=6.5cm,clip]{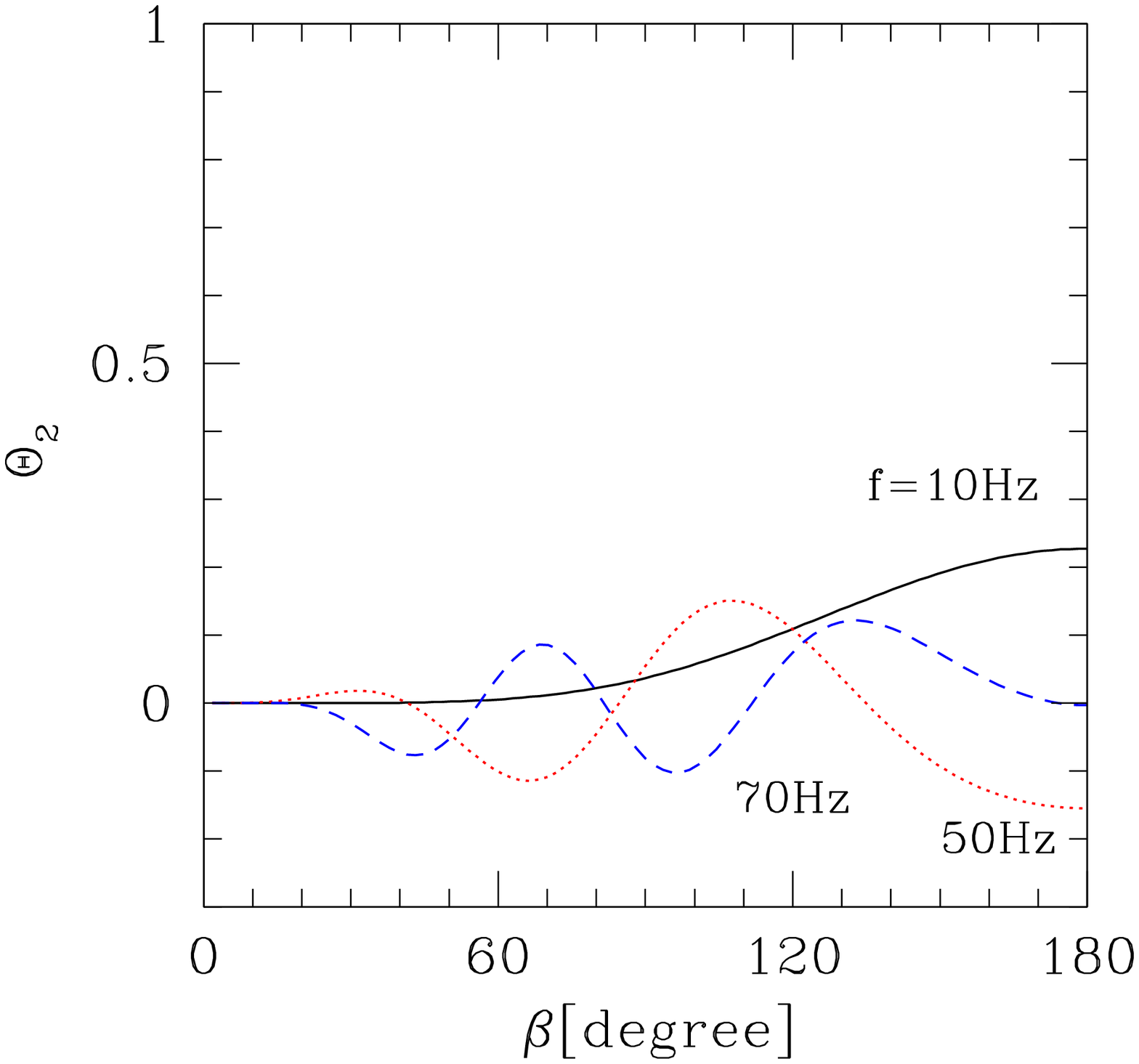}
\end{center}

\vspace*{-0.2cm}

\caption{The functions $\Theta_1(f,\beta)$ and  $\Theta_2(f,\beta)$  
for detectors on the
 Earth at frequencies $f=10$Hz, 50Hz and 70Hz.  }
\label{t1}
\begin{center}
\includegraphics[width=6.5cm,clip]{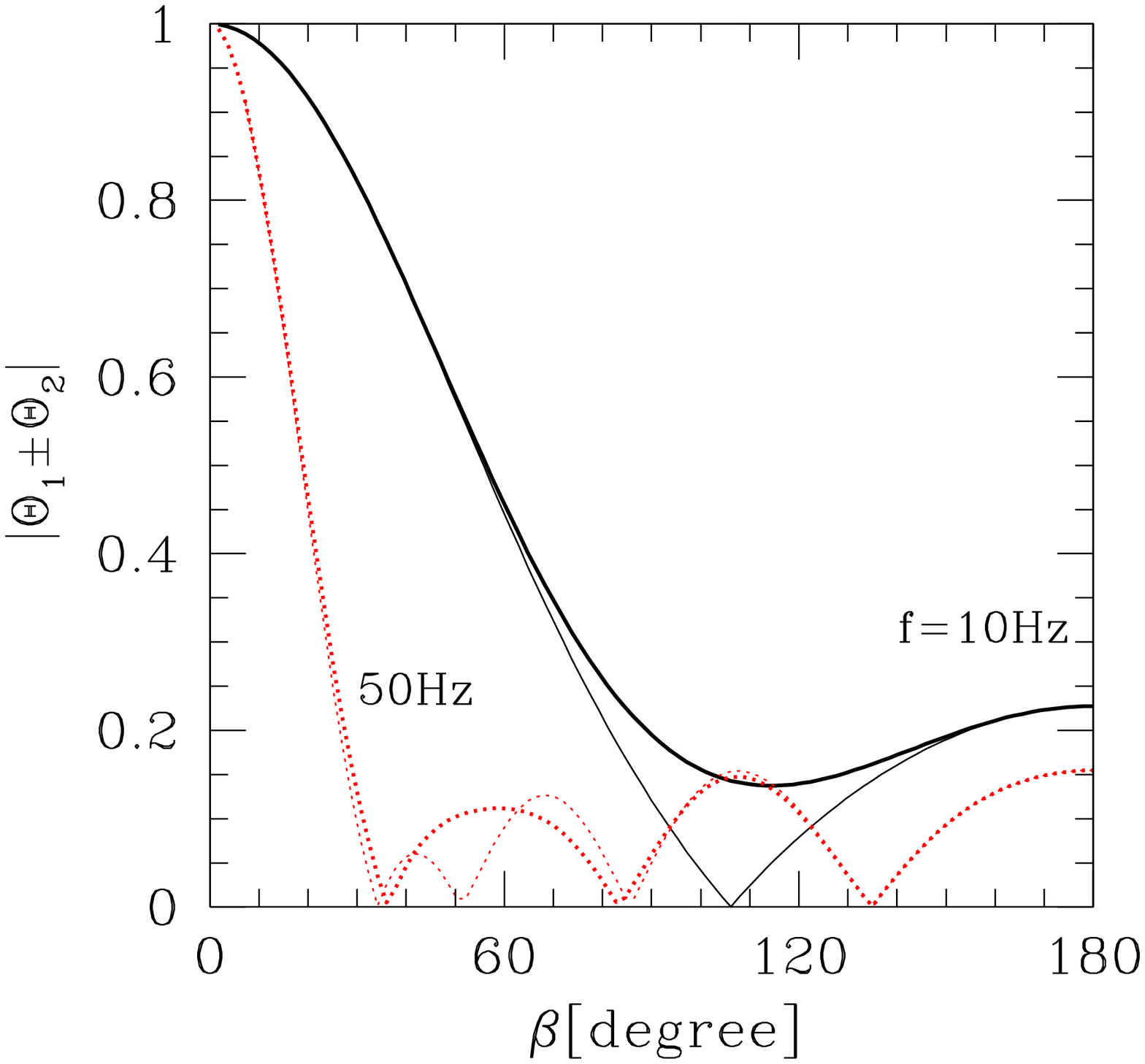}
\end{center}

\vspace*{-0.2cm}

\caption{The optimal combinations $|\Theta_1(f,\beta)+\Theta_1(f,\beta)|$
 (type II: thick lines) and $|\Theta_1(f,\beta)-\Theta_1(f,\beta)|$
 (type I: thin lines) for detectors on the
 Earth at frequencies $f=10$Hz and  50Hz.  }
\label{t12}
\end{figure}

In left panel of Figure \ref{t1}, the function $\Theta_1(f,\beta)$ 
is plotted against 
the angle parameter $\beta$ at specific frequencies $f=10$, 50 and 70Hz. 
As shown in Sec. \ref{subsec:same_plane}, 
we have $\Theta_1=1$ at $\beta=0^\circ$ that is 
the maximum value for $\gamma_I$ for given frequency $f$.
At frequency $f\ge 10$Hz relevant for ground-based detectors, the
function $|\Theta_1|$ becomes very small for a separation angle $\beta
\gsim 90^\circ$, and we 
identically have $\Theta_1=0$ at antipodal configuration $\beta=180^\circ$.
In right panel of Figure \ref{t1},  the shape of the second function 
$\Theta_2(f,\beta)$ is shown. As discussed in Sec. \ref{subsec:same_plane}, 
the function $\Theta_2$ becomes vanishing at $\beta=0^\circ$.  
This function exhibits an oscillatory behavior in the range
$0^\circ\le \beta \le  180^\circ$, and the number of its nodes 
is approximately 
proportional to $f R_{\rm E}$ (see appendix \ref{sec:tensor_analysis}).

In Figure \ref{t12}, we plot the overlap function $|\gamma_I|$ for two
optimal configurations, types I and II, at specific frequencies $10$ and 
$50$Hz.  The thin lines are for the 
type I with $|\gamma_I|=|\Theta_1-\Theta_2|$, while the
thick lines are for the type II with $|\gamma_I|=|\Theta_1+\Theta_2|$.
For angles close to 
$\beta=0^\circ$ and $180^\circ$, two lines are almost identical. 
This is because only one component is dominant there, that is, 
$|\Theta_1| \gg |\Theta_2| $ at $\beta\sim 0^\circ$, and  $|\Theta_2| \gg
|\Theta_1| $ at $\beta\sim 180^\circ$. Two components have comparable
magnitude at $\beta \sim 120^\circ$ for $f=10$Hz and at $\beta \sim 60^\circ$ 
for $f=50$Hz. As shown with the curves for $f=50$Hz, both types I and II
have chance to give the maximum value of $|\gamma_I|$ for given
$(f,\beta)$, depending on the relative signs of $\Theta_1$ and
$\Theta_2$.

\begin{figure}[t]
\begin{center}
\includegraphics[width=6.5cm,clip]{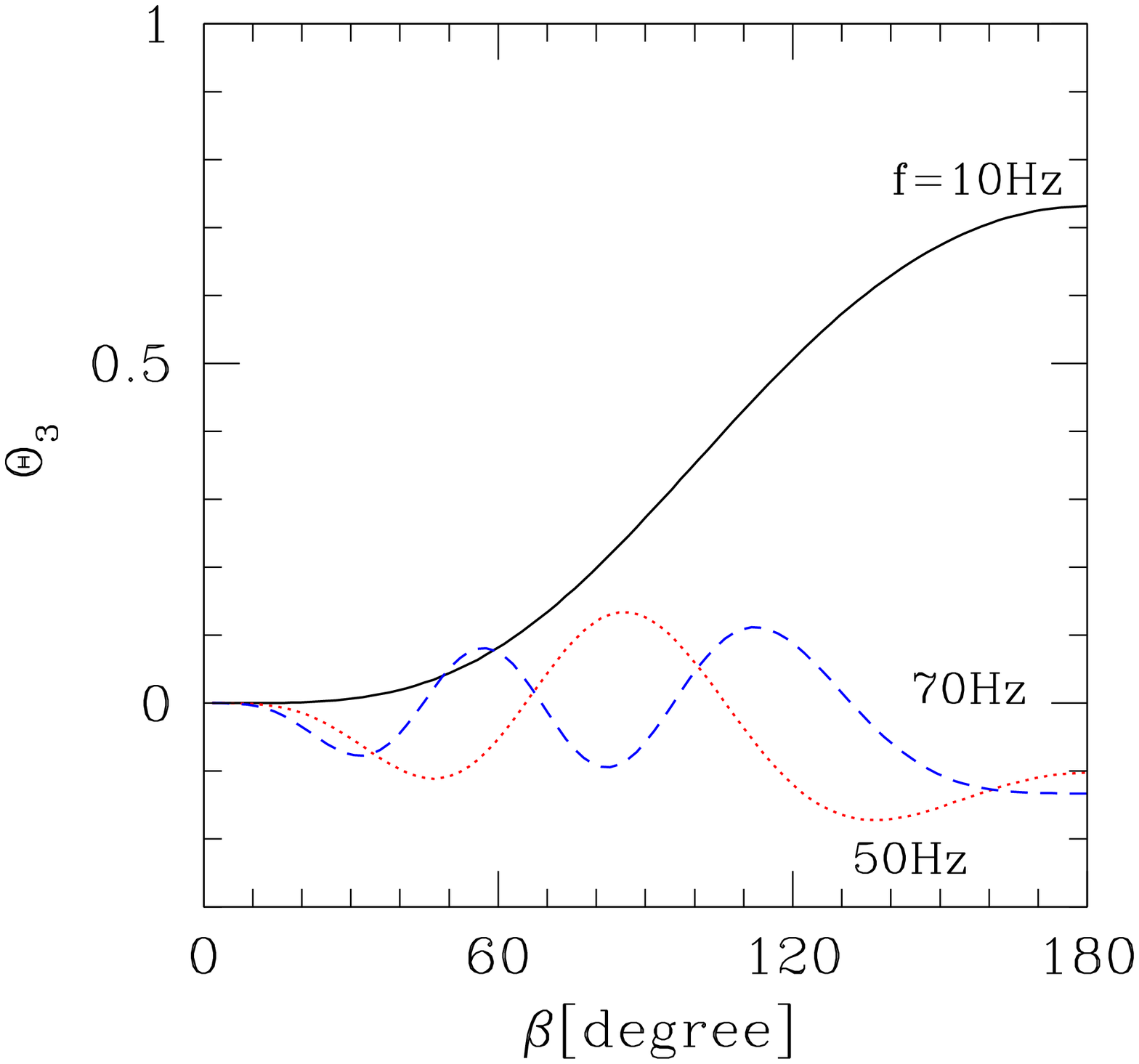}
\end{center}

\vspace*{-0.2cm}

\caption{The function $\Theta_3(f,\beta)$ for detectors on the
 Earth at frequencies $f=10$Hz, 50Hz and 70Hz.  }
\label{t3}
\begin{center}
\epsfxsize=6.5cm
\epsffile{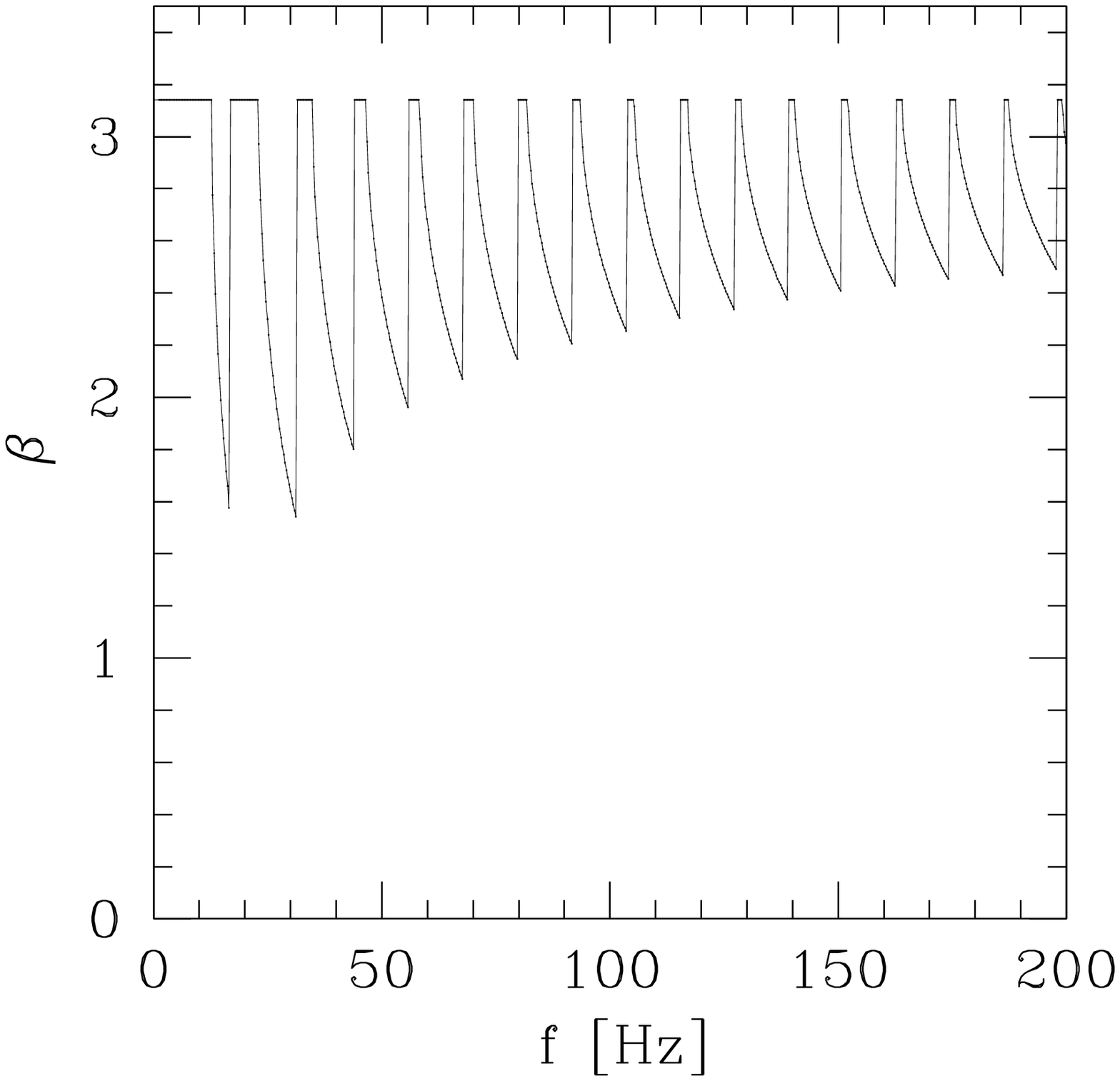}
 \end{center}

\vspace*{-0.2cm}

\caption{Values of $\beta_{\rm max}$ for detectors on the Earth 
as a function of frequency. While
 vertical lines are shown due to a software reason, there are
 discontinuities from $\beta_{\rm max}<180^\circ$ to 
 $\beta_{\rm max}=180^\circ$.  }
\label{beta}
\end{figure}
\begin{figure}[tb]
\begin{center}
\epsfxsize=6.5cm
\epsffile{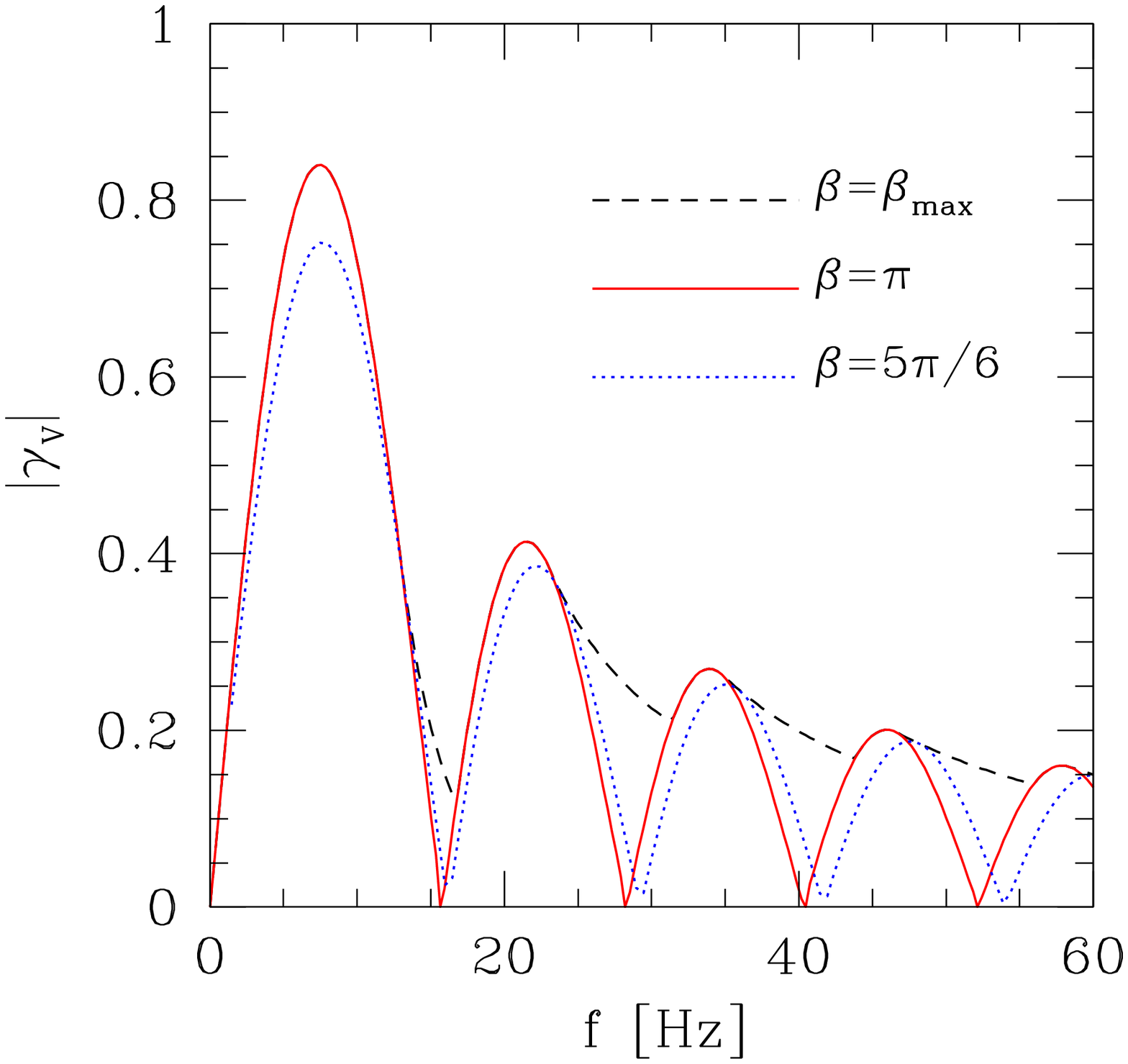}
 \end{center}

\vspace*{-0.2cm}

\caption{The function $|\gamma_V|$ for detectors on the Earth with type
 III configuration. The solid
 curve (dotted curve) is the result with $\beta=\pi$
 ($\beta=5\pi/6$). The dashed line is result with $\beta_{max}$ for
 which the function   $|\gamma_V|$ becomes maximum with given frequency
 $f$.  } 
\label{gvmax}
\end{figure}
Next, in Figure \ref{t3}, the function $\Theta_3$ for the $V$-mode 
is plotted. Note that we have $|\gamma_V|=|\Theta_3|$ for the type III 
configuration. As in the case for $\Theta_2$, the oscillating profiles have
a number of nodes approximately proportional to $fR_{\rm E}$. 
For given frequency $f$, we define  the 
separation angle $\beta_{\rm max}$ that maximizes the function  $|\Theta_3|$ in
the range $\beta \in [0^\circ,\,\,180^\circ]$.  In contrast to the simple 
results for the $I$-mode with 
$\max \gamma_I(f,\beta)=\Theta_1(f,\beta=0^\circ)=1$, the
angle $\beta_{\rm max}$ defined for the $V$-mode is slightly complicated 
and it does depend on the frequency $f$.
Figure \ref{beta} shows the angle $\beta_{max}$ in unit of radian, 
plotted against the frequency. The frequency dependence in the range 
$0<f<16.7$Hz can be understood with the following three steps:

\begin{description}
\item[(i)] As commented earlier, we have $\p_\beta 
\Theta_3(f,\beta=180^\circ)=0$ 
representing that the end point $\beta=180^\circ$ is generally an extreme.
At low frequency  regime,  the oscillating feature of $\Theta_3$ is
relatively simple (see Fig.~\ref{t3}), and the end point
$\Theta_3(f,\beta=180^\circ)$ is the global maxima.  We find 
$\beta_{\rm max}=180^\circ$ for $f\le 12.8$Hz.  
\item[(ii)] 
At $f=12.8$Hz, the end point $\beta=180^\circ$ becomes an inflection 
point with $\p^2_\beta \Theta_3(f,\beta=180^\circ)=0$.  Then,  
for $f>12.8$Hz,  there appears a local
maxima for $\Theta_3$ at $\beta<180^\circ$ that determines the separation
angle $\beta_{\rm max}$, 
as in Figure \ref{beta}. Meanwhile the end point 
$\Theta_3(f,\beta=180^\circ)$ is now a local minimum. With increasing $f$ it
decreases and crosses  0 at $f=15.7$Hz. 
\item[(iii)] The local maxima $\Theta_3(f,\beta_{\rm max})$  at 
$\beta_{\rm max}<180^\circ$
coincides with $-\Theta_3(f,\beta=180^\circ)$ ($>0$) at $f=16.7$Hz,
and the separation angle $\beta_{\rm max}$ shows a discontinuous 
transition up to $\beta_{\rm max}=180^\circ$ at $f=16.7$Hz. 
\end{description}

We can observe  similar cycles for the angle $\beta_{\rm max}$ at $f>16.7$Hz. 
The frequency dependent angle $\beta_{\rm max}$ should be regarded as the 
optimal separation for narrow band detection for the $V$-mode signal.
In Figure \ref{gvmax}, we show the maximum value
$|\Theta_3(f,\beta_{\rm max})|$ as well as $|\Theta_3(f,\beta=180^\circ)|$ 
and $|\Theta_3(f,\beta=150^\circ)|$. The  choice $\beta=150^\circ$ is just 
for an example. The first two curves coincide at some frequency bands 
(as shown in Fig. \ref{beta} for $\beta_{\rm max}$),
while the example  $|\Theta_3(f,\beta=150^\circ)|$ contacts with the dashed
curve for maximum value
$|\Theta_3(f, \beta_{\rm max}|$ only at specific  discrete frequencies.  In
the two dimensional region with 
$0^\circ\le \beta \le 180^\circ$ and $f\ge 0$, 
the global maximum  for $\Theta_3$ is 
\beq
-\frac{5}{32} (2\cos2-5\sin 2)=0.84,
\eeq
which appear at $\beta=180^\circ$ and $f=c/2\pi R_{\rm E}=7.5$Hz for
detectors on the Earth with $R_{\rm E}=6400$km.
In general, the function $\Theta_3$ is maximized at antipodal configuration 
($\beta=180^\circ$) with $f={c}/{2\pi R_{\rm s}}$, or equivalently 
$y=2$ due to the scaling commented in the beginning of this 
subsection. Although, for practical purpose to detect the $V$-mode signal, 
the broad-band analysis with multiple detectors is essential, 
which we will discuss in next section, 
it is clear from Figure \ref{gvmax} that the
separation $\beta=180^\circ$ seems the best choice for detectors, whose
bandwidths are much larger than the individual wiggle structure in this
figure.

\subsection{Overlap functions for specific pairs of detectors }
\label{subsec:overlap_specific}

Now, we analyze the geometry of ten pairs made from the five detectors 
listed in Table \ref{fig:f1}.  In this paper, we do not consider 
the co-located and co-aligned pair of detectors, such as two interferometers 
(4km+2km) at LIGO-Hanford. Co-located and co-aligned detectors are 
possibly contaminated by the measurement noises which are statistically 
correlated with each other, making the detection of stochastic signals 
difficult.

\begin{figure}[!tb]
\begin{center}
\epsfxsize=14cm
\epsffile{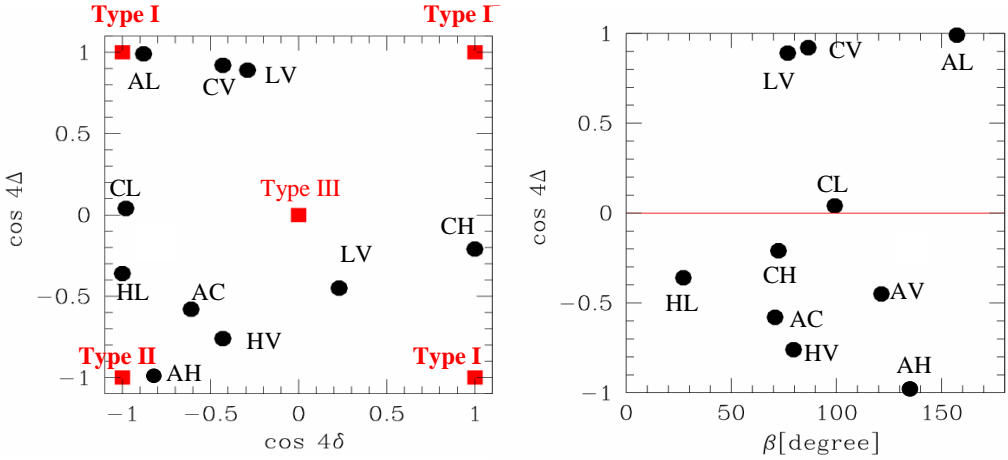}
\end{center}

\vspace*{-1.5cm}

\caption{{\it Left}: Distribution of the combinations 
  $(\cos 4\delta, \cos 4\Delta)$ for ten detector pairs shown in 
  Table \ref{tab:table2}.  Points for three types of configurations are 
  also given with 
  squares.  {\it Right}: Distribution of the 
  combinations $(\beta, \cos (4\Delta))$ for ten pairs shown in 
  Table \ref{tab:table2}. At relatively 
  large $\beta$ the sensitivity to the $V$-mode is roughly proportional 
  to $\sin (4\Delta)$. } 
\label{beta-del}
\end{figure}

Let us first examine how well the pairs of existing or planned 
interferometers are suitable for $I$- and $V$-mode detection 
by comparing the angle parameters with those of the 
optimal configuration discussed in Sec.~\ref{subsec:optimal_config}.  
In left panel of Figure~\ref{beta-del}, 
we plot the combination $(\cos (4\delta),\cos(4\Delta))$. 
>From this plot, the AL and AH pairs are 
found to be very close to the type I and type II configurations, 
respectively. Except for these, however, there are no other 
noticeable pairs. Turn to next consider a large separation angle 
$\beta\sim 180^\circ$, where the parameter $\delta$ becomes unimportant  
and the correlation signal can be approximately described by 
\beq
C_{ab}\simeq \frac{8\pi}{5}
\lkk\Theta_2(y,\beta)\cos(4\Delta)+\Theta_3(y,\beta)\sin(4\Delta) 
   \rkk. \label{largebeta}
\eeq
Thus, in this case, the angle parameters $\beta$ 
and $\Delta$ now play an important role. 
Since the regime $\beta \gsim 90^\circ$ is preferable for the $V$-mode 
detection, we next plot the combination $(\beta,\cos(4\Delta))$ 
in right panel of Figure~\ref{beta-del}. Among various pairs of detectors, 
the CL pair realizes nearly ideal angle ($\sin (4\Delta)=-1$), although the 
separation angle of CL pair is intermediate, i.e., $\beta=99.2^\circ$. 
Other interesting pairs for the $V$ mode with relatively large 
$|\sin (4\Delta)|$ are AV ($\sin (4\Delta)=-0.89$) and 
CH ($\sin(4\Delta)=0.98$).  
The HL pair has $\sin (4\Delta)=-0.93$, but its
separation is small, $\beta=27.2^\circ$, where the amplitude
$\Theta_3(y,\beta)$ is relatively small.

\begin{figure}[ht]
\begin{center}
\epsfxsize=8.9cm
\epsffile{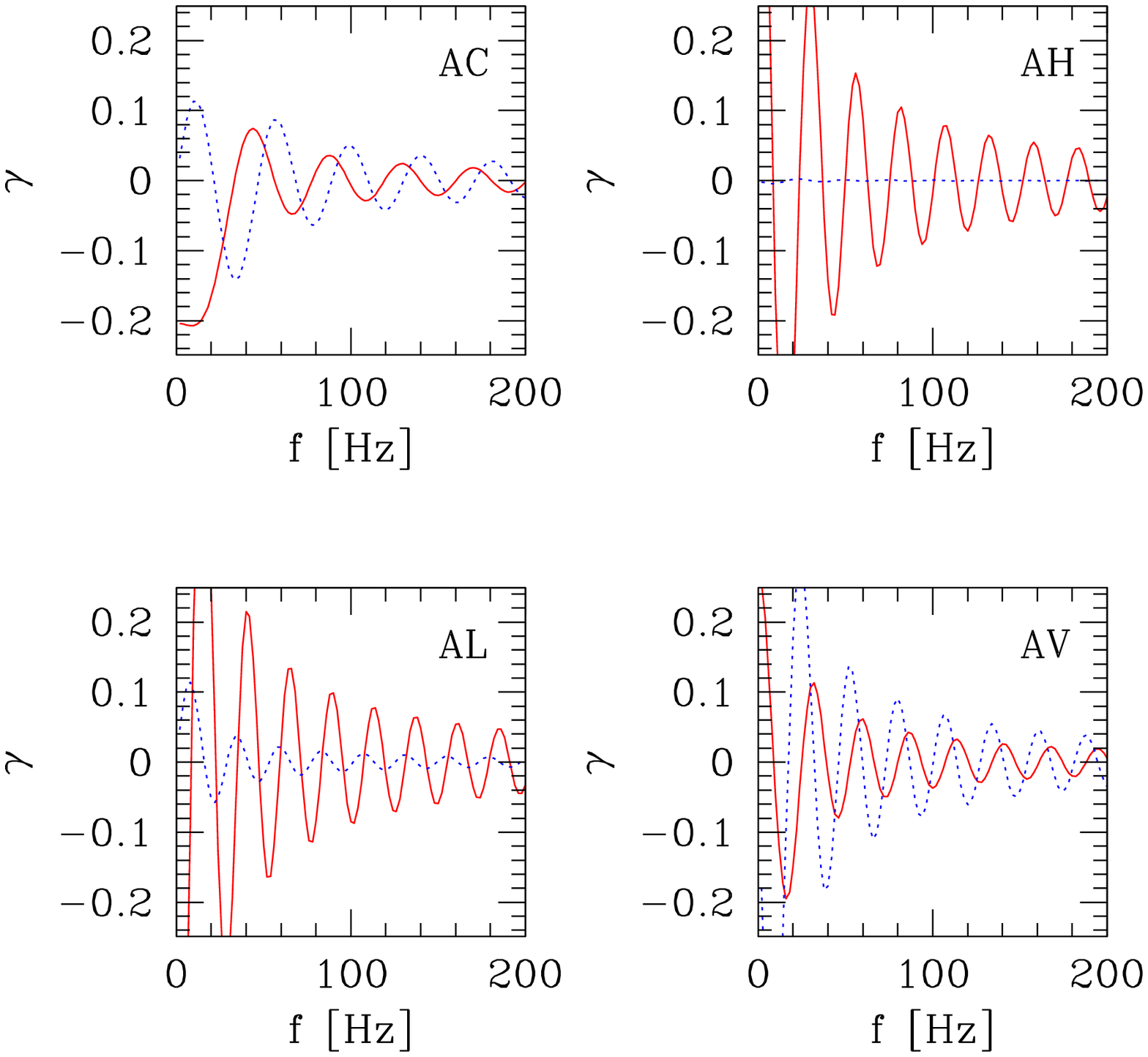}
\epsfxsize=8.9cm
\epsffile{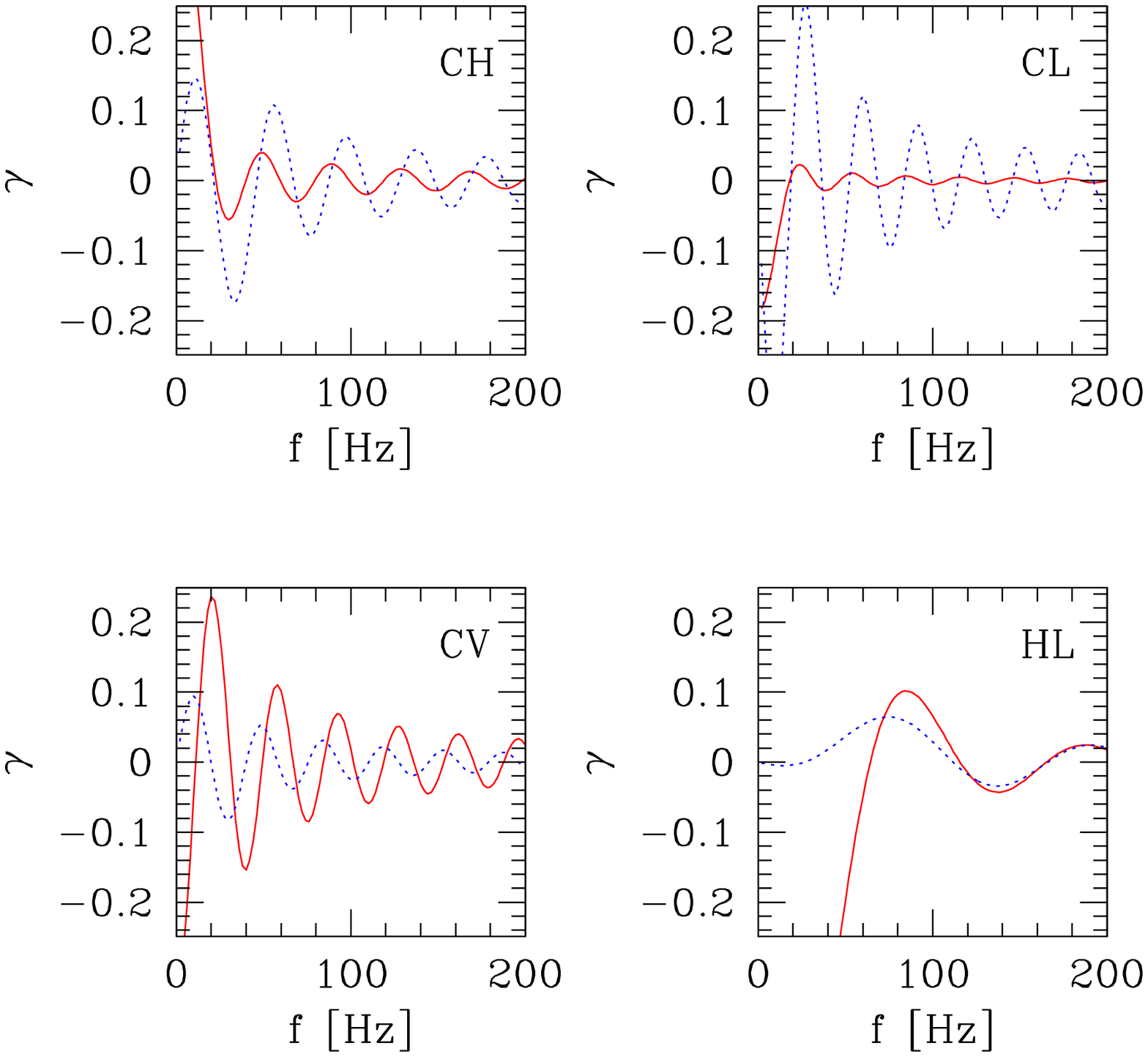}

\vspace*{0.7cm}

\epsfxsize=8.9cm
\epsffile{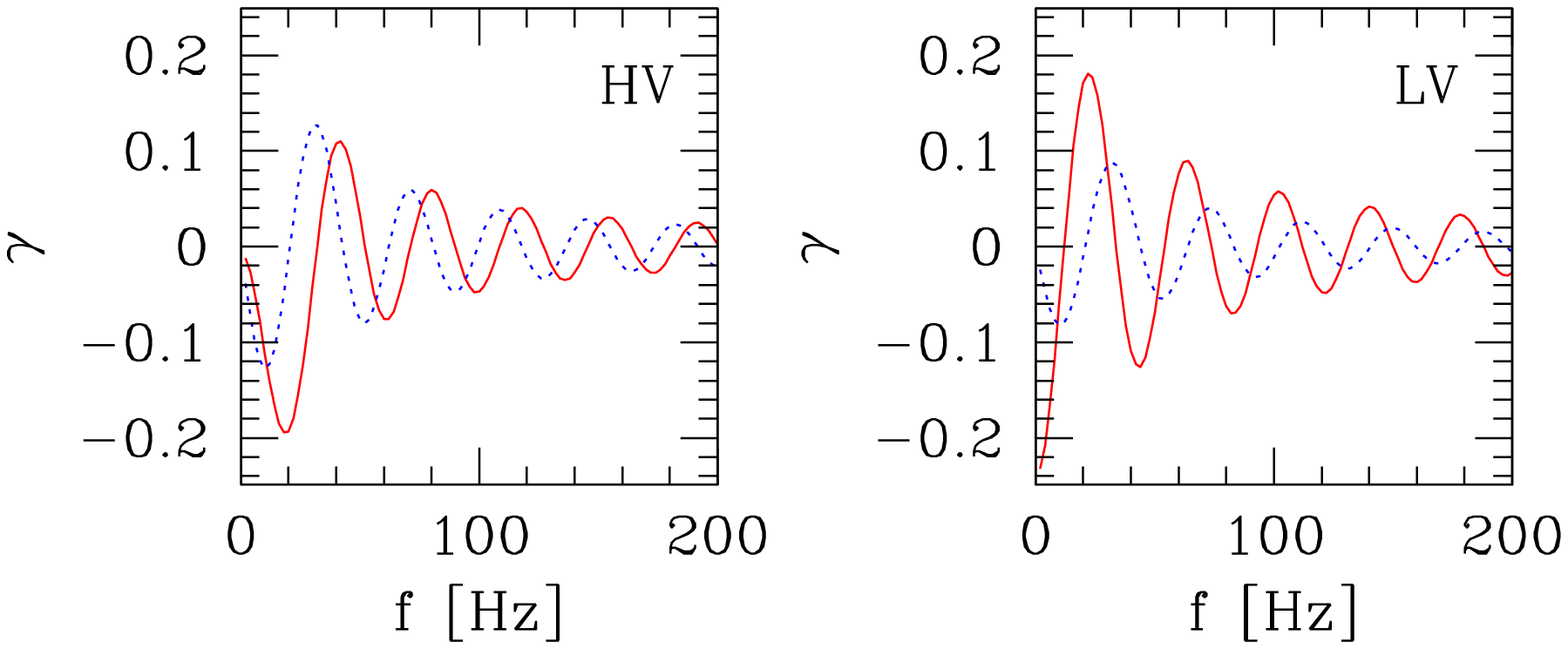}
\end{center}
\caption{Overlap functions for specific pair of detector made from the 
five detectors, A, C, H, L and V. The solid lines
 are for $\gamma_I$ and the dotted lines for $\gamma_V$. } 
\label{r1}
\end{figure}

Based on these considerations, 
in Figure \ref{r1}, we compile the overlap functions
($\gamma_I,\gamma_V$) for the ten pairs of detectors.  
There is characteristic frequency-width,  
$\Delta f\propto (\sin \beta/2)^{-1}$, determined by 
the arrival-time difference of gravitational waves between two cites.  
The frequency 
interval is largest for the HL pair.  For high frequencies, the peaks for the 
functions $\gamma_I$ and $\gamma_V$ have 1/4-cycle phase difference, as 
discussed in appendix \ref{sec:tensor_analysis}.  
The AH pair is almost insensitive to the $V$ mode, because it is 
close to the type II configuration.  The situation is similar for the AL
pair.  Note that the CL and AV pairs have relatively good sensitivity
to the $V$ mode, as anticipated from the angular parameters.  

\section{Broadband signal analysis }
\label{sec:broadband_SNR}

\subsection{Preliminary}
\label{subsec:preliminary}

So far, we have only dealt with the correlation signal of 
gravitational-wave backgrounds. In practice, the signal is 
contaminated by detector's noises,
 and thus the broadband signal analysis is essential ingredient 
for detection of background signals with high signal-to-noise ratio.

We model the data stream $s_a$ of a detector $a$ by 
a summation of gravitational-wave signal $H_a$ and detector noise $n_a$,
\beq
s_a=H_a+n_a.
\eeq
Throughout this paper, we assume that the noise of detector, $n_a$, obeys 
stationary and random processes and the noise correlation between 
any pair of detectors can be safely neglected. Then, 
covariance of the detector noises can be expressed as 
\beq
\lla n_a(f) n_b(f')^*  \rra=\frac12 \delta_{ab} 
\delta_{\rm D} (f-f')N_a(f), 
\eeq
where $N_a$ is the noise spectral density for detector $a$.

To estimate the sensitivity of each pair of detectors, 
let us consider the simple case with correlation 
$C_{ab}(f)$ of two detectors $a$ and $b$.  As it has been 
shown in the literature,  
the total signal-to-noise ratio (SNR) is given by 
\cite{Flanagan:1993ix,Allen:1997ad} (see also appendix 
\ref{sec:PDF_for_corr})
\beqa
\mathrm{SNR}^2&=&\lmk\frac{16\pi}{5}  \rmk^2 T_{\rm obs} 
\lkk 2  \int_0^\infty df
\frac{(\gamma_{I} I+\gamma_{V} V)^2}{ {\cal N}_{ab}(f)}
\rkk
\nonumber
\\ 
&=&\lmk\frac{3H_0^2}{10\pi^2}  \rmk^2 T_{\rm obs} \lkk  2\int_0^\infty df
\frac{\Omega_{\rm GW}(f)^2 (\gamma_{I}+
\gamma_{V}\Pi)^2}{f^6 {\cal N}_{ab}(f)}
\rkk  \label{broad}
\eeqa
with the quantity ${\cal N}_{ab}$ defined by 
${\cal N}_{ab}\equiv N_a(f)N_b(f)$. This is the result obtained 
in the weak-signal limit $I(f)\ll N_{\{a,b\}}(f)$. Note that 
the above formula just represents the SNR for the total amplitude of 
the background signals, and it does not imply the SNR for a pure $I$- or 
$V$-mode signal. The separation of $I$- and $V$-modes will be 
discussed in next section.

For quantitative evaluation of SNRs,  we need an explicit 
form of the noise spectral density. In the following, 
we use the fitting form of the noise spectra for 
advanced LIGO detector, $N_{\rm ligo}$. 
Assuming that all the detectors have identical 
noise spectra with $N_{\rm ligo}$, signal-to-noise ratios of stochastic 
signals are estimated. Based on Ref.\cite{adv},  
the analytical fit of the noise spectrum $N_{\rm ligo}$ is given by 
\beq
N_{\rm ligo}(f)= 
\left\{
\begin{array}{lcl}
{\displaystyle 10^{-44} \lmk \frac{f }{\rm 10 Hz}\rmk^{-4}+10^{-47.25 }
\lmk \frac{f }{\rm 100 Hz} \rmk^{-1.7}}  {\rm Hz^{-1}} & {\rm for} &
10{\rm Hz}\le f \le 240{\rm Hz},
\\
{\displaystyle 10^{-46} \lmk \frac{f }{\rm 1000 Hz} \rmk^{3} } 
 {\rm Hz^{-1}} & {\rm for} &  240{\rm Hz}\le f \le 3000{\rm Hz}, 
\\
\infty &~& {\rm otherwise}.
\end{array}
\right.
\eeq

\begin{figure}[!tb]
\begin{center}
\epsfxsize=10.cm
\epsffile{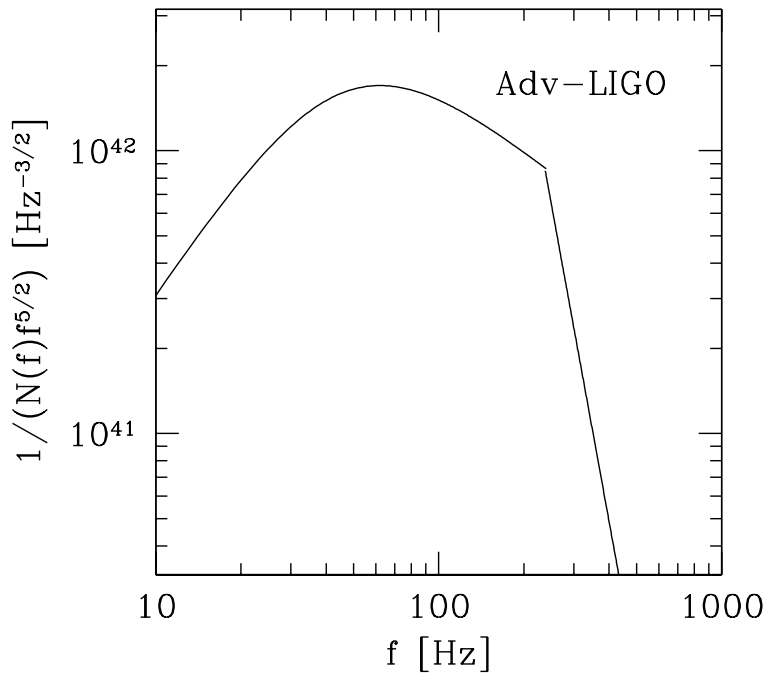} 
\end{center}

\vspace*{-0.2cm}

\caption{ The weight function $(N(f)f^{5/2})$ for advanced LIGO.  } 
\label{nc}
\end{figure}

The expression (\ref{broad}) implies that the weight function for 
SNR per logarithmic frequency interval $d\ln f$ is 
proportional to $(N(f)f^{5/2})^{-1}$ for flat input $\Omega_{\rm GW}(f) 
(\gamma_{I}+\gamma_{V}\Pi)={\rm const}$. In Figure \ref{nc}, using 
the analytic form of the noise spectrum, we plot the weight function. 
It becomes maximum around $\sim 50$Hz with its bandwidth $\sim 100$Hz. 
Note that the shape of the weight function for stochastic signals is close 
to the one for the signals produced by binary neutron stars, in which case  
the detectable distance, as the integral of weight function, is 
roughly proportional to 
\beq
\lkk \int_0^\infty d\ln f \frac{1}{f^{4/3} N(f)}   \rkk^{1/2}.
\eeq
The next-generation detectors are primarily designed to have the 
good sensitivity to a chirping signal of binary neutron stars, 
and they are planned to achieve the similar performance for detecting these binaries. 
In this sense,   
our assumption that all the detectors have identical 
noise spectrum with advanced LIGO is reasonable.

Finally, as a reference, we present the SNR for coincident
detectors ($\gamma_{I,ab}=1, \gamma_{V,ab}=0$):      
\beq
\mathrm{SNR}_0=4.8\lmk \frac{T_{\rm obs}}{3\rm yr}  \rmk^{1/2}\lmk
\frac{\Omega_{\rm GW}h_{70}^2}{10^{-9}}  \rmk.  \label{norm}
\eeq
This value will be frequently referred, as a baseline of the SNR for 
various situations considered below.

\subsection{Signal-to-noise ratios for  pair of detector}
\label{subsec:SNR}

The total SNR (\ref{broad}) depends strongly on model parameters of the
background, including the polarization degree $\Pi$. 
In order to present our numerical results concisely,  we use the
normalized form
\beq
Q\equiv\frac{\rm SNR}{\rm SNR_0}.
\eeq
To characterize sensitivity of each pair to $I$- or $V$-mode signal, 
based on the above equation, we respectively define  $Q_I$ and $Q_V$ by 
replacing $\{I,\,\,V\}$ in the expression of SNR 
with $\rho_c/(4\pi^2f^3)\{\Omega_{\rm GW},\,\,0\}$ 
and $\rho_c/(4\pi^2f^3)\{0,\,\,\Omega_{\rm GW}\}$. 
These normalized SNRs  $Q_I$ and $Q_V$ can be regarded as a rms value 
of overlap functions with the weight function $(f^{5/2} N(f))^{-1}$.

\begin{figure}[!bth]
\begin{center}
\epsfxsize=9.0cm
\epsffile{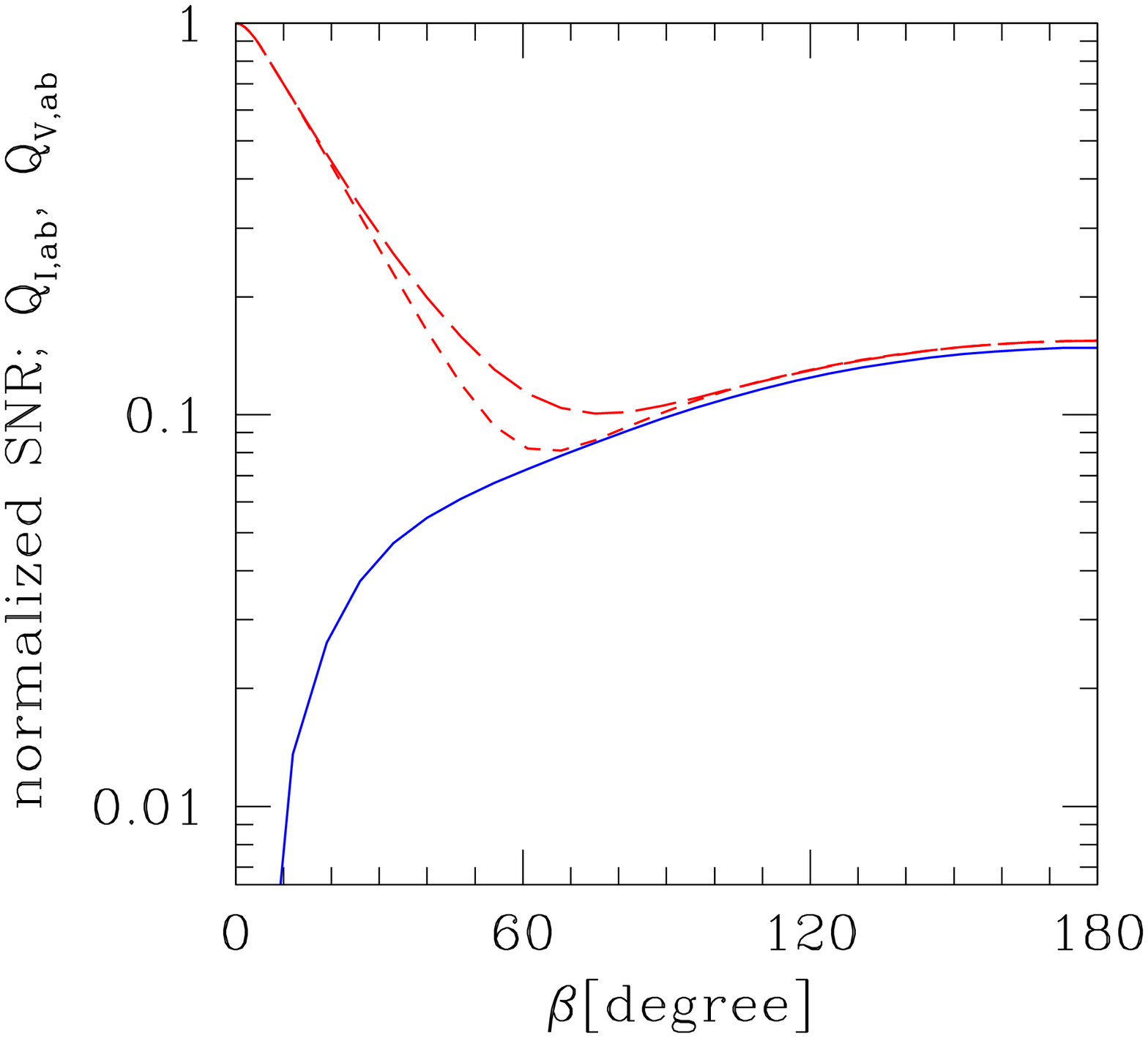}
 \end{center}

\vspace*{-0.2cm}

\caption{ Normalized signal to noise ratios (${ Q}_{I,ab}$ and 
  ${ Q}_{V,ab}$) with optimal configurations  
for the $I$-mode (short dashed curve: type I, long dashed curve: type II)
 and for  the $V$-mode (solid curve: type III with setting $\Pi=1$ for
 illustrative purpose). We use the noise curve for
 the advanced LIGO.  } 
\label{nsr}
\begin{center}
\epsfxsize=5.8cm
\epsffile{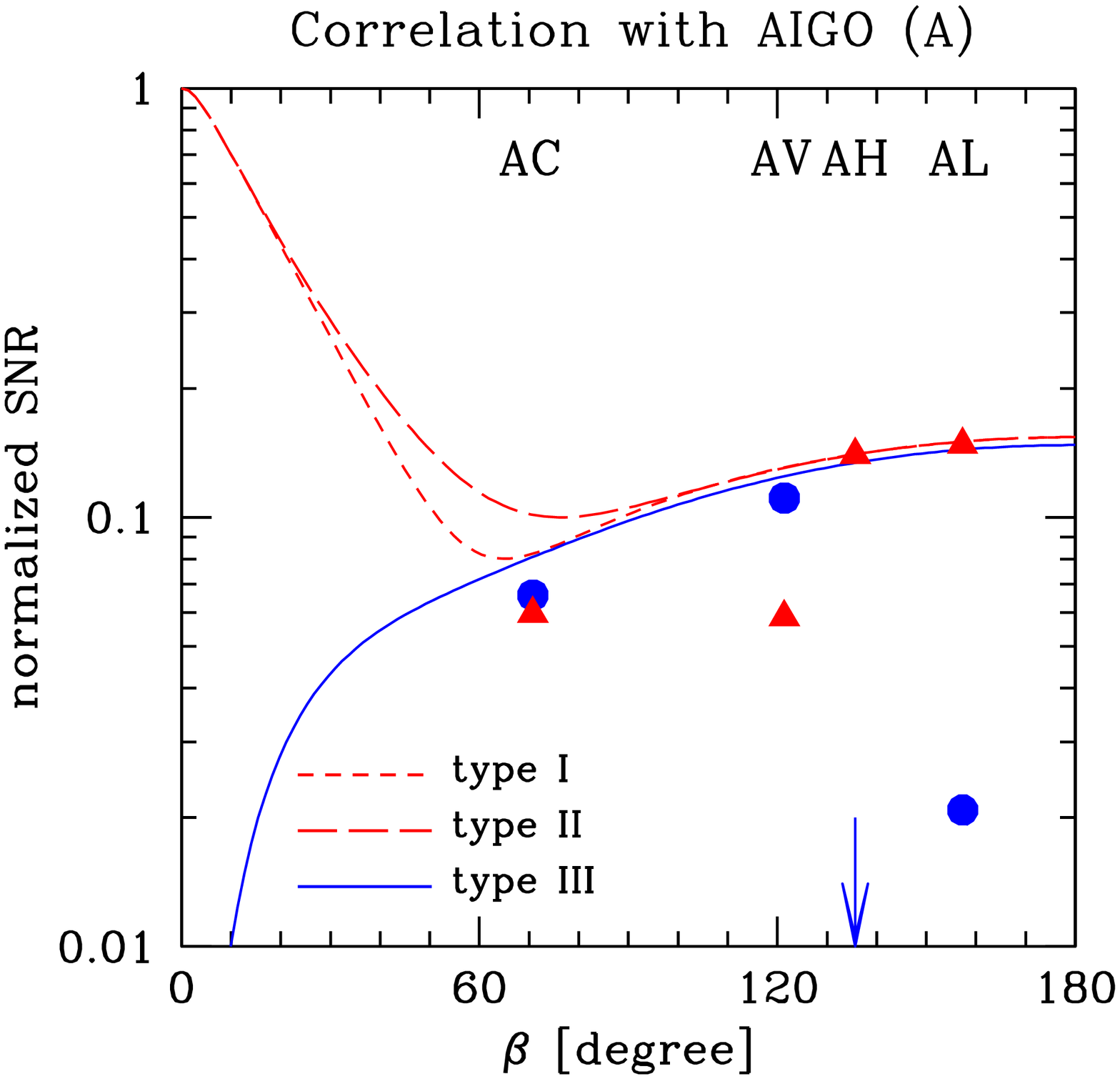}
\epsfxsize=5.8cm
\epsffile{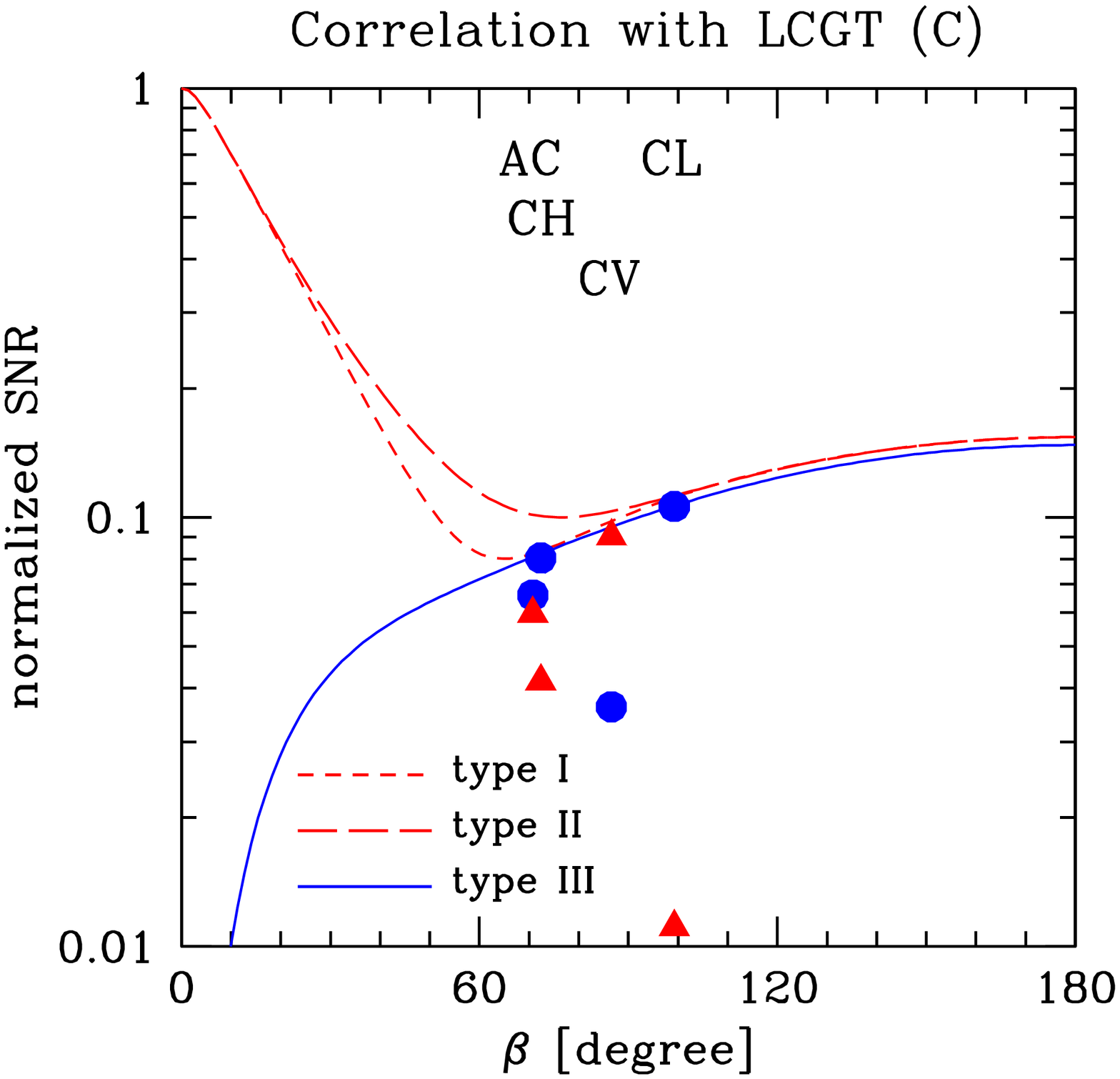}
\epsfxsize=5.8cm
\epsffile{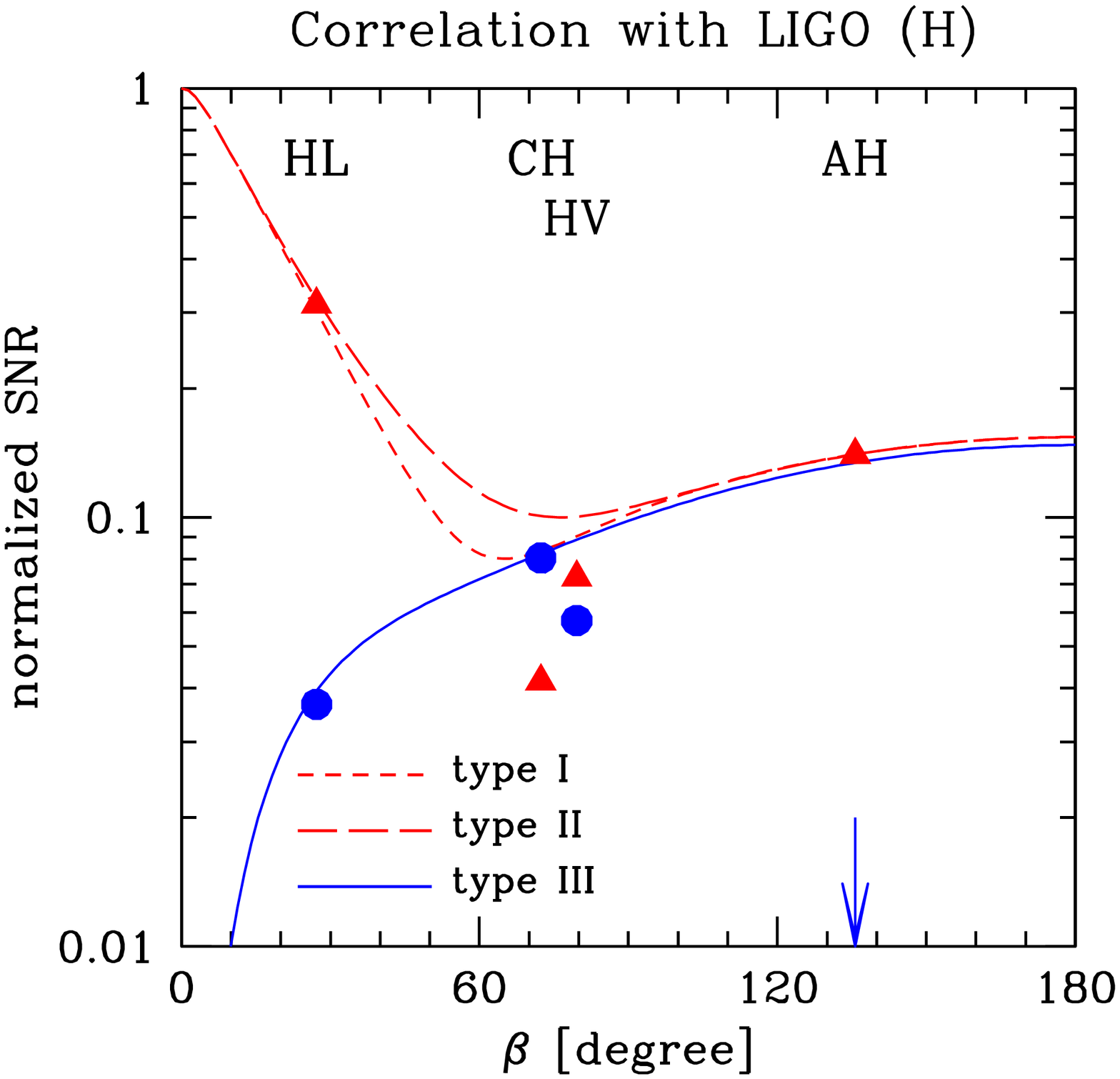}
\epsfxsize=5.8cm
\epsffile{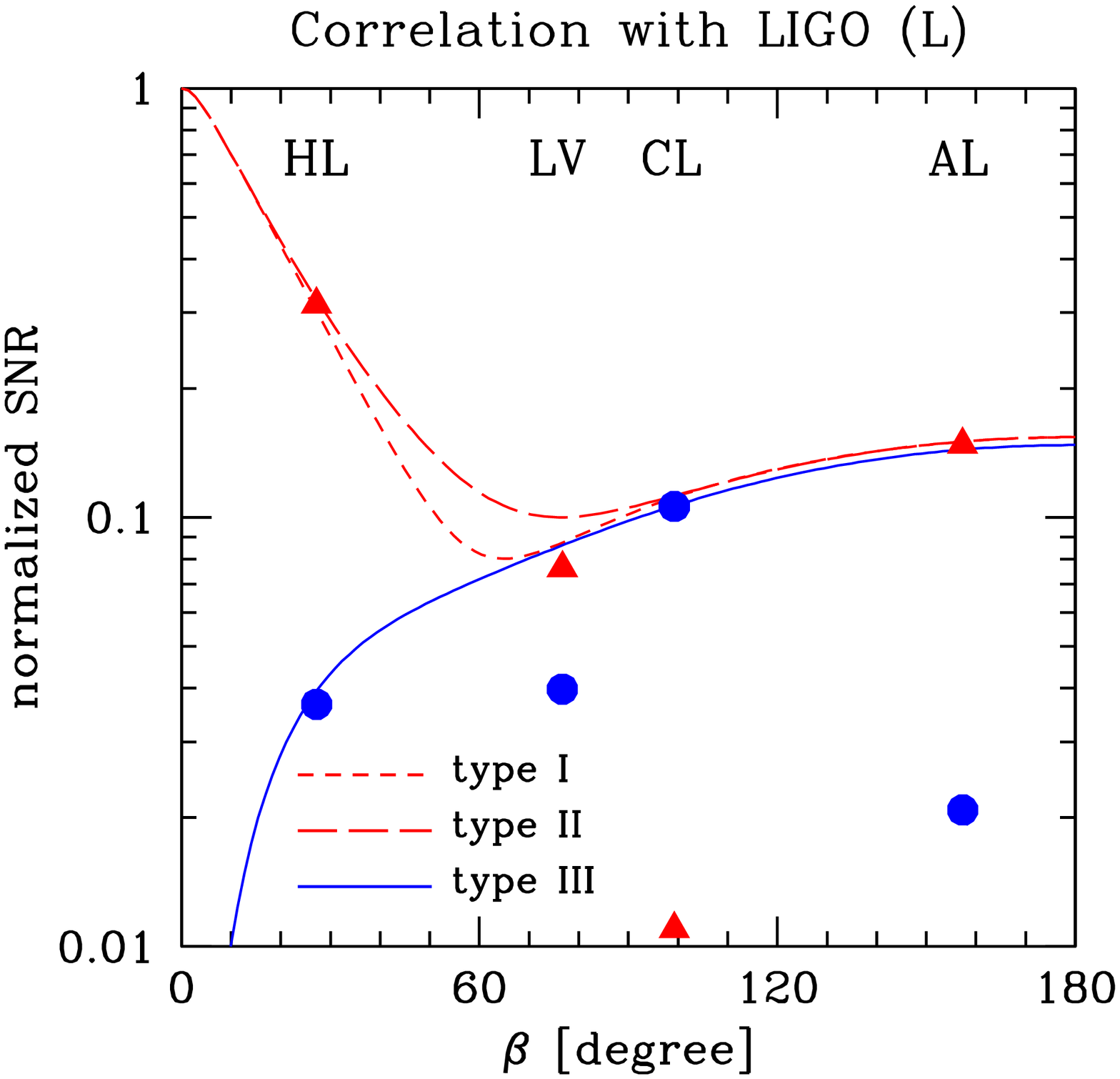}
\epsfxsize=5.8cm
\epsffile{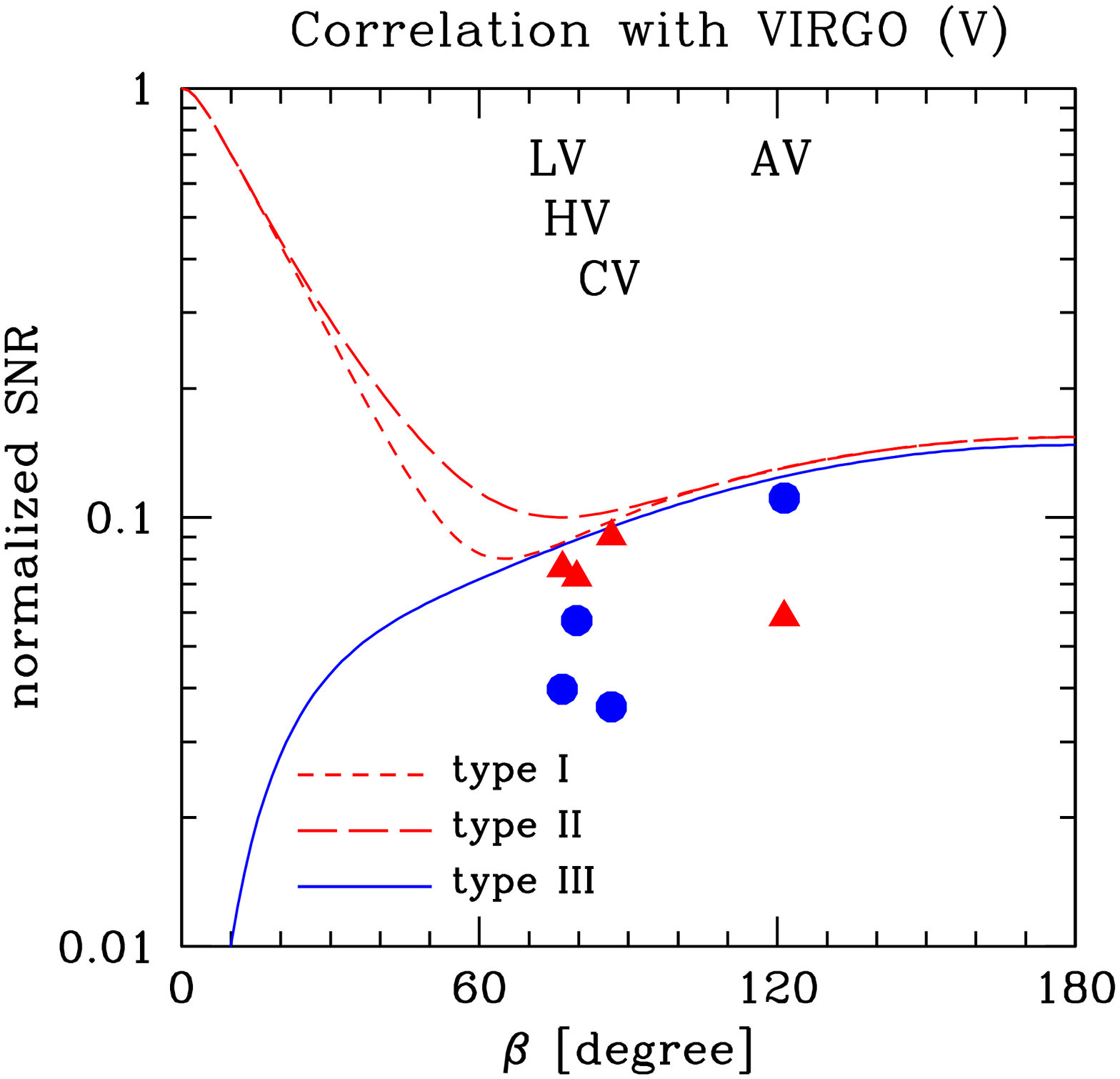}
 \end{center}

\vspace*{-0.2cm}

\caption{ The normalized SNRs $Q_I$ (circles) and $Q_V$ 
(triangles) for detector pairs.} 
\label{net}
\end{figure}

In Figure \ref{nsr}, we present the normalized SNR 
for the optimal geometry, i.e., types I, II and III configurations 
(short-dashed, long-dashed and solid lines, respectively). 
One noticeable point is that a widely separated ($\beta\sim 180^\circ$) 
pair is powerful to 
search for the $V$ mode. At $\beta=180^\circ$, we find the following 
asymptotic relations
\beq
\gamma_I\sim-\frac{5}{2}\cos(4\Delta) \frac{\sin y}{y},
\quad
\gamma_V\sim\frac{5}{2}\sin(4\Delta) \frac{\cos y}{y}
\eeq
with $y=4\pi R f/c$.  For detectors on the Earth, the characteristic 
frequency interval is $c/2R_{\rm E}\sim20$Hz, which is enough inside the 
bandwidth of advance LIGO, $\Delta f\sim 100$Hz. As a result, the 
oscillation of the overlap functions are averaged out and the normalized 
SNRs $Q_I$ and $Q_V$ give a similar output for optimal configurations 
(types I-III) at $\beta=180^\circ$.

In Figure \ref{net} and Table \ref{tab:Q_I_Q_V},  we show the 
normalized SNRs for pairs made from the five  interferometers in 
Table \ref{tab:location_of_detectors}. 
To reduce the contribution from the $I$-mode signal, 
pairs that have been regarded as disadvantageous for constraining 
$\Omega_{\rm GW}$, can now play important roles for measuring the $V$ mode, 
according to equation (\ref{largebeta}). 
The HL pair with $\cos (4\delta)\sim 1$ and $\sin (4\Delta)\sim 0.93$ 
realizes nearly maximum values simultaneously for ${ Q}_{I,ab}$ and 
${ Q}_{V,ab}$, at its separation $\beta=27.2^\circ$. 
This is because ${Q}_{I,ab}$ is mainly determined by the angle $\delta$ 
at a small  $\beta$, while  ${Q}_{V,ab}$ depends only on
$\Delta$. This pair has the largest $Q_I$ among ten pairs of
detectors. In contrast, the CL pair has good sensitivity to $V$, 
although it is relatively insensitive to the $I$ mode, because of 
$\sin (4\Delta)\sim 1$ and  $|\cos(4\Delta)|=0.04$. 
Indeed, the orientation of the LCGT detector is only $1.2^\circ$ different 
from the optimal direction ($\cos(4\Delta)=0$) with respect to the 
LIGO-Livingston cite. As other interesting pairs, the AH pair is almost 
insensitive to the $V$ mode with $\sin(4\Delta)=-0.007$ 
(nearly type II configuration with LIGO-Hanford). 
The AV pair has a large $Q_V$  with
$|\sin (4\Delta)|=0.89$, but its $Q_I$ is much larger than 
that of the CL pair. 
\begin{table}[!tbh]
\begin{tabular}{l|c|c|c|c|c}
\hline\hline
~ & ~~A~~  & ~~C~~  & ~~H~~  & ~~L~~  & ~~V~~ \\ 
\hline
~~A~~& $*$  &  $0.060$ &  $0.14$  &  $0.15$  &  $0.059$ \\ 
\hline
~~C~~ & $0.07$  &  $*$  &  $0.042$  & $0.011$  & $0.091$ \\ 
\hline
~~H~~ & $0.0009$  & $0.081$ &  $*$  & $0.32$ & $0.073$ \\ 
\hline
~~L~~ &  $0.021$ & $0.11$  & $0.037$ & $*$  & $0.077$ \\ 
\hline
~~V~~ & $0.11$  & $0.036$ & $0.058$ & $0.040$ & $*$ \\ 
\hline\hline
\end{tabular}
\caption{Normalized SNRs $Q_I$ (upper-right) and $Q_V$ (lower-left).
}
\label{tab:Q_I_Q_V}
\end{table}

\subsection{Antipodal detectors}
\label{subsec:antipodal}

So far, we have studied pairs of detectors on the Earth,  
strictly keeping the radius of sphere $R_{\rm E}=6400$km,    
for which the antipodal type III configuration is turned out to be 
optimal to realize the largest SNR $Q_V$.  Here, we discuss to what 
extent one can improve the sensitivity to the $V$ mode by varying the 
radius of sphere.

\begin{figure}[!tb]
\begin{center}
\epsfxsize=9.cm
\epsffile{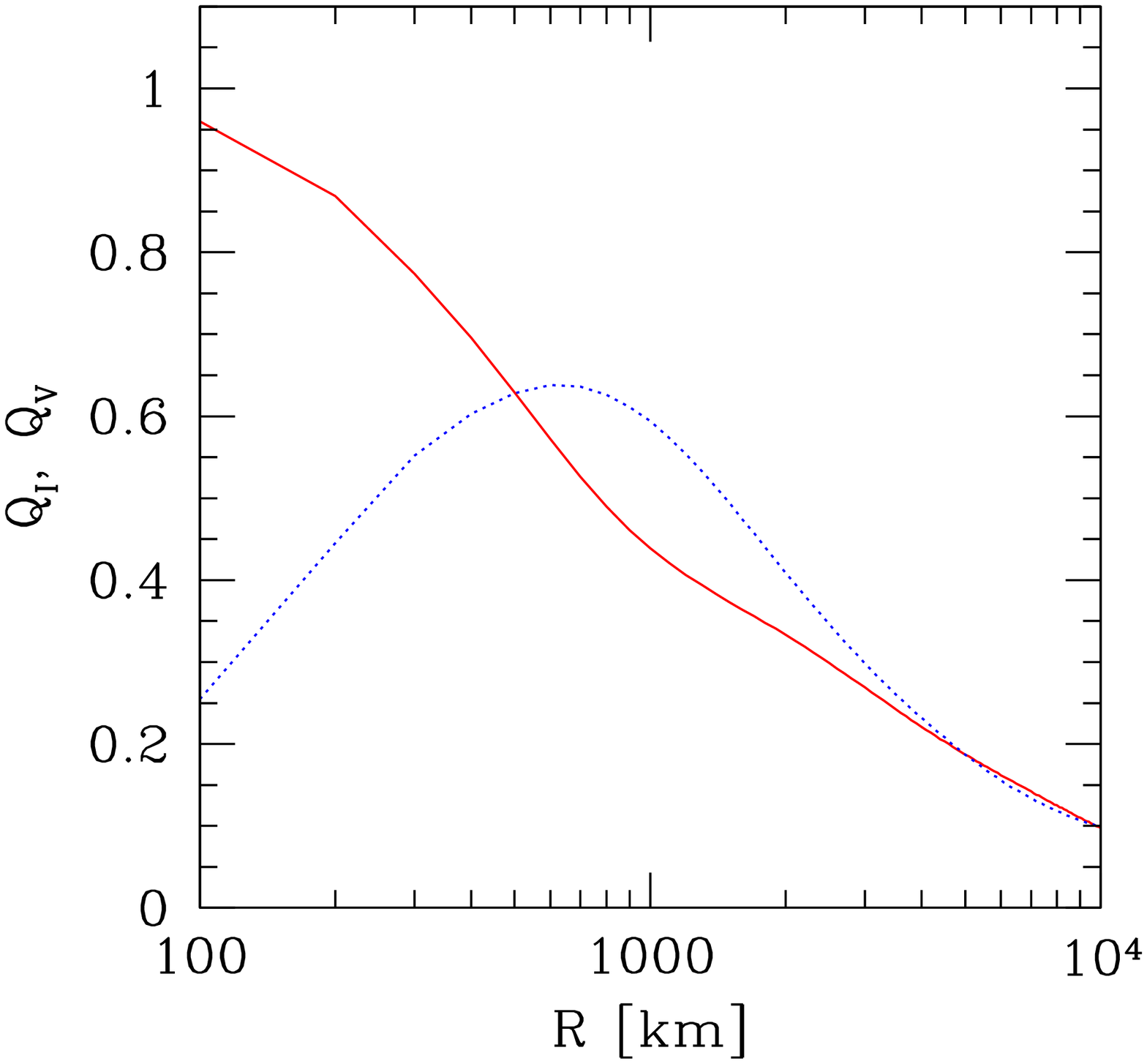}
\end{center}

\vspace*{-0.2cm}

\caption{ The normalized SNR for antipodal configuration with radius
 $R$.  Advanced LIGO noise curve is used. } 
\label{anti1}

\vspace*{0.5cm}

\begin{center}
\epsfxsize=14.cm
\epsffile{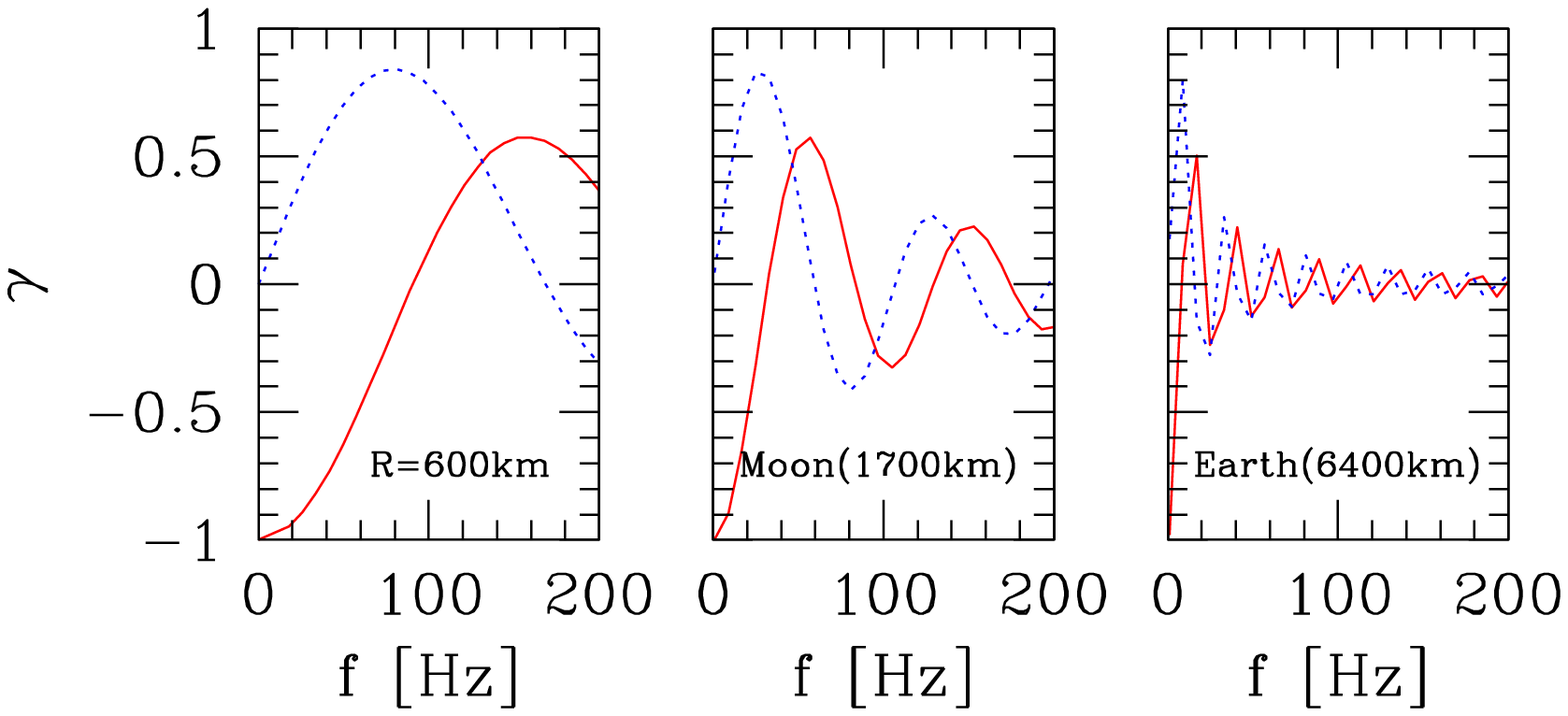}
\end{center}

\vspace*{-0.2cm}

\caption{ Overlap functions for antipodal detector pairs (solid curves:
 $\gamma_I$, dotted curves: $\gamma_V$). } 
\label{anti2}
\end{figure}

In Figure \ref{anti1}, we plot the broadband sensitivities $Q_I$ and 
$Q_V$ for antipodal detectors as functions of the radius of sphere, $R$. 
Note that the noise spectrum is fixed to $N_{\rm ligo}(f)$ as before. 
Here, we put $\cos (4\Delta)=1$ for $Q_I$ and 
$\sin (4\Delta)=1$ for $Q_V$.  While the value $Q_I$ is maximized at 
$R=0$ and we obtain $Q_I=1$ and $Q_V=0$, the maximum value of $Q_V$ is 
achieved when $R=600$km, leading to $Q_V=0.64$.  
It is interesting to note that at $R\sim 1700$km corresponding to the 
radius of the Moon, we still obtain rather larger value, $Q_V=0.45$.

In Figure \ref{anti2}, the overlap functions for three representative cases 
are plotted: $R=600$, $1700$ and $6400$km. As we commented earlier, 
the overlap functions for any radius can be rescaled and become 
identical if we plot the functions against the rescaled variable, 
$y\propto fR$. Further, for the configuration examined in Figure 
\ref{anti1}, the relations $\gamma_I(0)=-1$ and $\gamma_V(0)=0$ strictly 
hold. In those situations, the shape of the noise spectrum 
shown in Figure \ref{nc} is the key to determine the best value for 
$Q_V$ and the overlap function $\gamma_V$ for $R\sim 600$km 
eventually becomes the best shape to achieve the maximum value of  
$Q_V=0.64$. Also, it turned out that the sensitivity for the 
Moon becomes about three times larger than that for the Earth, $Q_V=0.15$ (see appendix \ref{sec:moon} for comments on detectors on the Moon).

\section{Separation}
\label{sec:separation}

As discussed so far, a simple analysis with correlation signal 
$C_{ab}$ of two detectors allows us to detect a mixture of 
$I$- and $V$-mode signals, but we cannot extract each of them 
separately. In order to disentangle these two signals, 
in this section, we discuss the problem of the $I$- and $V$-mode 
separation considered in Ref.\cite{Seto:2007tn}. After  
describing the simplest case using a set of four detectors 
in Sec.\ref{subsec:two_correlation}, we generalize it to 
the multiple-detector case in Sec.~\ref{subsec:multiple_correlation}, 
and present a statistical framework to achieve the optimal sensitivity.  
Based on these theoretical backgrounds, in Sec.~\ref{subsec:network},  
the correlation analysis with network of ground-based detectors are examined 
and the optimal values of SNR are derived for each $I$ and $V$ mode.

\subsection{Analysis with two correlation signals}
\label{subsec:two_correlation}

Let us begin by considering the simplest case that two pairs 
of interferometers $(a,b)$ and $(c,d)$ are available. We write 
down their correlation signals as
\beq
C_{ab}(f)=\frac{8\pi}5\{\gamma_{I,ab}(f) I(f)+\gamma_{V,ab}(f) V(f)\},
\quad
C_{cd}(f)=\frac{8\pi}5\{\gamma_{I,cd}(f) I(f)+\gamma_{V,cd}(f) V(f)\}.
\label{2eq}
\eeq
Making their linear combination, $I$ or $V$ mode can be 
removed, and a pure $V$ or $I$ mode is separately extracted. 
Except for trivial scaling, the unique choices are
\beq
{\gamma_{V,ab}C_{cd}-\gamma_{V,cd}C_{ab}}=
\frac{8\pi}5 I{(\gamma_{I,cd}\gamma_{V,ab}-\gamma_{I,ab}\gamma_{V,cd})},
\quad
{\gamma_{I,ab}C_{cd}-\gamma_{I,cd}C_{ab}}=
\frac{8\pi}5 V{(\gamma_{V,cd}\gamma_{I,ab}-\gamma_{V,ab}\gamma_{I,cd})}.
\eeq
Meanwhile, the noise spectra for these combinations are proportional to 
\beq
{\gamma_{V,ab}^2 {\cal N}_{cd}+\gamma_{V,cd}^2 {\cal N}_{ab}},
\quad
{\gamma_{I,ab}^2 {\cal N}_{cd}+\gamma_{I,cd}^2 {\cal N}_{ab}},
\eeq
where quantity ${\cal N}_{ab}$ indicates the product of noise spectra, 
${\cal N}_{ab}=N_a(f)N_b(f)$. Taking account of proportional 
factors, the broadband signal-to-noise ratio for a pure $I$ or 
$V$ mode becomes
\beqa
\mathrm{SNR}_I^2&=&
\lmk\frac{16\pi}{5}\rmk^2 T_{\rm obs} \lkk  2\int_0^\infty df
\frac{I^2(\gamma_{V,cd}\gamma_{I,ab}-\gamma_{V,ab}\gamma_{I,cd})^2}
{ (\gamma_{V,ab}^2 {\cal N}_{cd}+\gamma_{V,cd}^2 {\cal N}_{ab})}  \rkk, 
\label{eq:compiled_SNR_I}
\\
\mathrm{SNR}_V^2&=&
\lmk\frac{16\pi}{5}\rmk
^2 T_{\rm obs} \lkk  2\int_0^\infty df
\frac{V^2(\gamma_{V,cd}\gamma_{I,ab}-\gamma_{V,ab}\gamma_{I,cd})^2}
{ (\gamma_{V,ab}^2 {\cal N}_{cd}+\gamma_{V,cd}^2 {\cal N}_{ab})}  \rkk. 
\label{eq:compiled_SNR_V}
\eeqa
For detectors with identical  noise spectra  with 
${\cal N}(f)=N(f)^2$, the {\it compiled} overlap functions are defined as
\beq
\Gamma_{I,ab:cd}\equiv
\frac{\gamma_{I,cd}\gamma_{V,ab}-\gamma_{I,ab}\gamma_{V,cd}}{[\gamma_{V,ab}^2 
+\gamma_{V,cd}^2]^{1/2}},
\quad
\Gamma_{V,ab:cd}\equiv
\frac{\gamma_{V,cd}\gamma_{I,ab}-\gamma_{V,ab}\gamma_{I,cd}}{[\gamma_{I,ab}^2 
+\gamma_{I,cd}^2]^{1/2}}.
\label{co}
\eeq
With these functions, the broadband SNRs for the separated two modes 
can be estimated from equation (\ref{broad}) just replacing  
the term $\gamma_II+\gamma_VV$ with $\Gamma_{I,ab:cd} I$ and 
$\Gamma_{V,ab:cd}V$.  In this sense, 
the complied overlap functions represent the sensitivities
to the $I$ and $V$ modes after the separation.

\begin{figure}[!tb]
\begin{center}
\epsfxsize=12cm
\epsffile{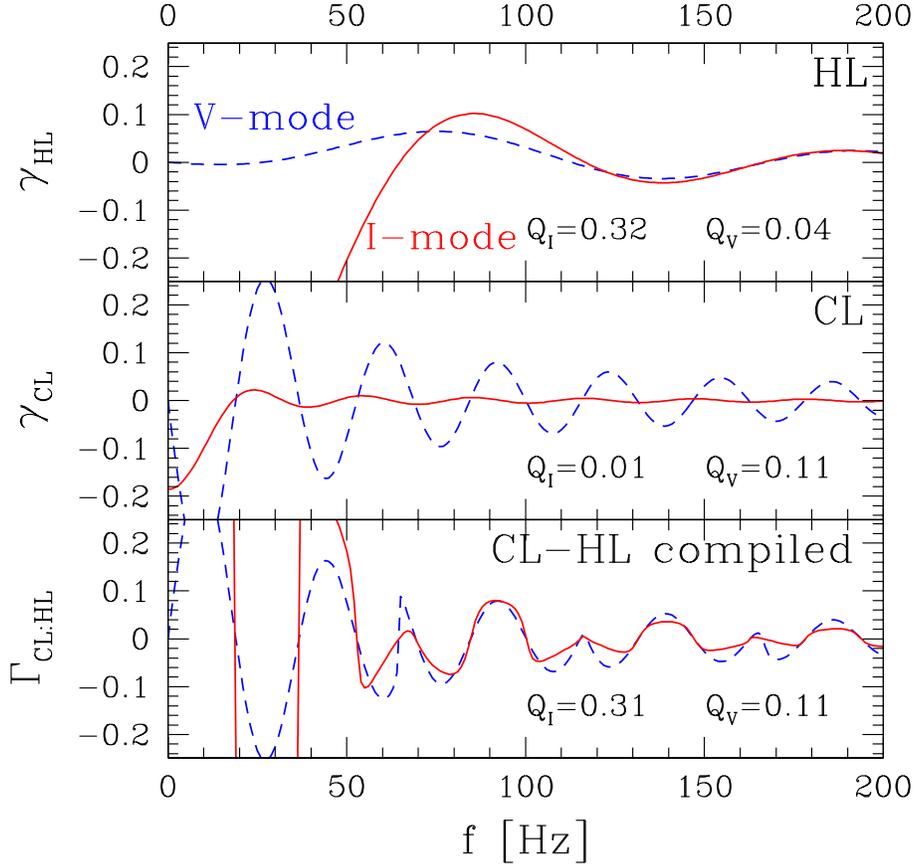}
 \end{center}

\vspace*{-0.3cm}

\caption{Overlap  functions for the unpolarized $I$ mode (dashed
 curves), and the circularly polarized $V$ mode (solid curves). The
 upper panel shows 
 the results for the Hanford-Livingston (HL) pair. The middle one
 is results for the LCGT-Livingston (CL) pair. The
 normalized SNRs ${\it S}_{I,V}$ (with the adv LIGO noise 
spectrum) are also presented.  The bottom one show the compiled
 functions $\Gamma_{I,V}$ (Eq.(\ref{co})) made from  both pairs.
  } 
\label{twodet}
\end{figure}

In Figure \ref{twodet}, as a specific example for the mode separation, 
we consider the HL and CL pairs and plot the compiled overlap function, 
as well as  the overlap functions for each pair. Although the 
shapes of the functions $\Gamma_{ab:cd}$ are very complicated, 
resultant values of the normalized SNR estimated from 
equations (\ref{eq:compiled_SNR_I}) and (\ref{eq:compiled_SNR_V}) 
are  $0.11$ for the $V$ mode and $0.31$ for the $I$ mode, which are 
very close to the values presented in Sec.~\ref{subsec:SNR}  
(${Q}_V=0.11$ for CL, ${Q}_I=0.32$ for HL). 
Note also that for other combinations, the $I$ and $V$-mode separation 
can be performed with nominal changes to the naively expected 
sensitivities ${Q}_{\{I,V\},ab}$. This will be discussed in 
details in Sec.~\ref{subsec:network}.

\subsection{Analysis with multiple data set }
\label{subsec:multiple_correlation}

Next consider the generalization of the previous analysis to the 
cases with multiple data set. 
For a network of $n$ detectors, we can make totally $n_t=n(n-1)/2$ 
independent correlation signals 
\beq
C_i=\frac{8\pi}5(\gamma_{Ii}\, I+\gamma_{Vi}\, V)
\eeq
with $i=1,.., n_t$.
Here, we use the single suffix $i$ to represent a detector pair for which
we have assigned two suffixes so far, such as $ab$  in equation (\ref{2eq}).
For the number of detectors with $n>2$, the number of output signals 
becomes $n_t>3$, and this implies that 
we must deal with the over-determined problem in order 
to separate a mixture of $I$- and $V$-modes, because 
the number of observables 
exceeds the number of target parameters, $I$ and $V$. 
In what follows, we will discuss the signal-to-noise ratios expected
from the optimal data analysis.

Let us examine a straightforward extension of the analysis in previous 
subsection. Provided the original data set of correlation signals, 
$\{C_i\}$, obtained from all possible pairs of detectors, we can make 
the linear combinations, $D_{Ii}$ and $D_{Vi}$, which respectively eliminate 
the variable $V$ and $I$:  
\beq
D_{Ii}=d_{Ii} I,~~~ D_{Vi}=d_{Vi} V,
\eeq
where $d_{Ii}$ and $d_{Vi}$ denote some numerical coefficients, 
appropriately chosen for removing the contribution from $V$ and $I$, 
respectively.  Note that the  number 
of independent combinations labeled as $i$ is 
 $n_t-1$ \footnote{One simple example is to make 
$D_{Ii}=C_i-({\gamma_{Vi}}/{\gamma_{Vn_t}})\,C_{n_t}=
(\gamma_{Ii}-\gamma_{In_t}\gamma_{Vi}/\gamma_{Vn_t})\,I$ 
and $D_{Vi}=C_i-({\gamma_{Ii}}/{\gamma_{In_t}})\,C_{n_t}=
(\gamma_{Vi}-\gamma_{Vn_t}\gamma_{Ii}/\gamma_{In_t})\,V$.}. 
The associated covariance matrices for intrinsic noises, 
${\cal M}^{-1}_{Iij}$ and ${\cal M}^{-1}_{Vij}$, are expressed
in terms of the quantities, 
$\gamma_{Ii}$, $\gamma_{Vi}$ and ${\cal N}_i$ (noise spectra).  
Then, the resultant total SNRs for the optimal combinations of 
data sets $\{D_{Ii},D_{Vi}\}$ are 
\beq
\mathrm{SNR}_I^2\propto d_{Ii}{\cal M}^{-1}_{}d_{Ij},
\quad
\mathrm{SNR}_V^2\propto d_{Vi}{\cal M}^{-1}_{Vij}d_{Vj}.
\label{eq:SNR_direct_method}
\eeq
For $n_t=n=3$, the above expressions can be recast in a 
rather simple form: 
\beqa
\mathrm{SNR}_I^2&\propto &
I^2\lkk \lmk \sum_i^3 \frac{\gamma_{Ii}^2}{{\cal N}_i}  \rmk  \lmk
\sum_i^3 \frac{\gamma_{Vi}^2}{{\cal N}_i} \rmk - \lmk \sum_i^3
\frac{\gamma_{Ii}\gamma_{Vi}}{{\cal N}_i}  \rmk^2  \rkk  \lmk \sum_i^3
\frac{\gamma_{Vi}^2}{{\cal N}_i}  \rmk^{-1},
\label{31}
\\
\mathrm{SNR}_V^2&\propto &
V^2\lkk \lmk \sum_i^3 \frac{\gamma_{Ii}^2}{{\cal N}_i}  \rmk  \lmk
\sum_i^3 \frac{\gamma_{Vi}^2}{{\cal N}_i} \rmk - \lmk \sum_i^3
\frac{\gamma_{Ii}\gamma_{Vi}}{{\cal N}_i}  \rmk^2  \rkk  \lmk \sum_i^3 
\frac{\gamma_{Ii}^2}{{\cal N}_i}  \rmk^{-1}.
\label{32}
\eeqa
Equations (\ref{31}) and (\ref{32}) are symmetric with respect to 
the suffix $i$ and they do not depend on the specific choice of 
the data sets $\{D_{Ii},D_{Vi}\}$. 
Self-consistently, the above expressions recover the 
previous results, (\ref{eq:compiled_SNR_I}) and (\ref{eq:compiled_SNR_V}), 
if we set $\gamma_{I3}$ and $\gamma_{V3}$ to zero. 
Indeed, the symmetric expressions (\ref{31}) and (\ref{32}) for $n_t=3$ 
generally hold for the cases with $n_t>3$ and we will use these 
forms to estimate the SNRs for optimal combination of five 
ground-based detectors. In appendix \ref{sec:derivation_optimal_SNR}, 
a brief sketch to derive the symmetric expressions for the $n_t>3$ cases 
is presented.  Multiplying the factor $2(16\pi/5)^2df$ and integrating 
over entire frequency range, 
the narrow band SNRs (\ref{31}) and (\ref{32}) can be generalized to 
the broadband SNRs
\beqa
\mathrm{SN
R}_I^2&= & 2\lmk \frac{16\pi}5 \rmk^2  T_{\rm obs} \int_0^\infty df 
I^2\lkk \lmk \sum_i^{n_t} \frac{\gamma_{Ii}^2}{{\cal N}_i}  \rmk  \lmk
\sum_i^{n_t} \frac{\gamma_{Vi}^2}{{\cal N}_i} \rmk - \lmk \sum_i^{n_t}
\frac{\gamma_{Ii}\gamma_{Vi}}{{\cal N}_i}  \rmk^2  \rkk  \lmk \sum_i^{n_t}
\frac{\gamma_{Vi}^2}{{\cal N}_i}  \rmk^{-1},
\label{312}
\\
\mathrm{SNR}_V^2&= & 2\lmk \frac{16\pi}5 \rmk^2  T_{\rm obs} \int_0^\infty df
V^2\lkk \lmk \sum_i^{n_t} \frac{\gamma_{Ii}^2}{{\cal N}_i}  \rmk  \lmk
\sum_i^{n_t} \frac{\gamma_{Vi}^2}{{\cal N}_i} \rmk - \lmk \sum_i^{n_t}
\frac{\gamma_{Ii}\gamma_{Vi}}{{\cal N}_i}  \rmk^2  \rkk  \lmk \sum_i^{n_t} 
\frac{\gamma_{Ii}^2}{{\cal N}_i}  \rmk^{-1}.
\label{322}
\eeqa

Similar to the one defined in previous subsection, 
we define the effective overlap functions 
for detectors with identical noise spectra, 
which represent the optimal sensitivities to the $I$ and $V$ modes: 
\beq
\Gamma_{{\rm eff},I}\equiv \lmk\frac{\sum_i^{n_t} \gamma_{iI}^2\sum_i^{n_t}
\gamma_{iV}^2-(\sum_i^{n_t} \gamma_{iI}\gamma_{iV})^2}{\sum_i^{n_t}
\gamma_{iV}^2}  \rmk^{1/2},
\quad
\Gamma_{{\rm eff},V}\equiv \lmk\frac{\sum_i^{n_t}
\gamma_{iI}^2\sum_i^{n_t} \gamma_{iV}^2-(\sum_i^{n_t}
\gamma_{iI}\gamma_{iV})^2}{\sum_i^{n_t} \gamma_{iI}^2}  \rmk^{1/2}.  
\eeq
For sensitivity only for the $I$ or $V$ mode,  we also define 
\beq
\Gamma_{{\rm eff},I0}=\lmk \sum_i^{n_t} \gamma_{iI}^2\rmk^{1/2},
\quad
\Gamma_{{\rm eff},V0}=\lmk \sum_i^{n_t} \gamma_{iV}^2\rmk^{1/2}, 
\eeq
which correspond to the effective overlap function (with identical 
noise spectrum) for the traditional analysis in the absence of $V$ or 
$I$ mode. In the following, we use the notation $Q_{I0}$ for the 
normalized SNR with effective function $\Gamma_{{\rm eff},I0}$. 
Based on these definitions,  the ratio $R$ is given by 
\beq
R= \frac{\sum_i^{n_t} \gamma_{iI}\gamma_{iV}}
{\Gamma_{{\rm eff},I0}\Gamma_{{\rm eff},V0}}.
\label{eq:def_R}
\eeq
As increasing the number of detectors, the functions
$\Gamma_{{\rm eff},I0}$ and $\Gamma_{{\rm eff},V0}$ monotonically 
increase. For a large numbers of detectors, however, the numerator
$\sum_i^{n_t} \gamma_{iI}\gamma_{iV}$ can be regarded as a summation 
of random numbers, and the ratio $R$ is expected to decrease quickly. 
We will see this numerically in next subsection.

\subsection{Optimal SNRs from ground-based network}
\label{subsec:network}

\begin{figure}[!bth]
\begin{center}
\epsfxsize=10.cm
\epsffile{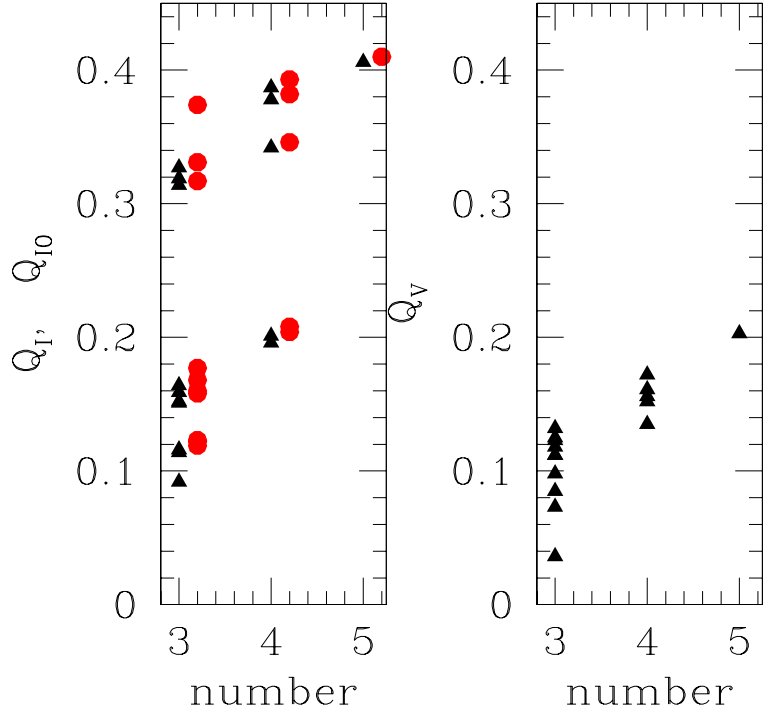}
 \end{center}
\caption{ The normalized SNRs $Q_I$ (left panel; filled triangles),
 $Q_{I0}$ (left panel ; open circles), and $Q_V$ (right panel; filled
 triangle).  The horizontal axis is the number of detectors.  We
 slightly shift the points for $Q_{I0}$ to the right.} 
\label{f25}
\end{figure}

We are in position to evaluate the broad band SNRs for optimal combination 
of network of five detectors, A,C,H,L and V.  
For networks made by three detectors among five detectors,  
there are  ${}_5C_3=10$ possible networks. 
In the same way, we can make ${}_5C_4=5$ networks for combinations of 
four detectors. Numerical results for detector networks are presented 
in Table \ref{nett}. In Figure \ref{f25}, we also provide the normalized 
SNRs for various combinations of detectors, showing the 
overall behaviors against the number of detectors.  
\begin{table}[!bth]
\begin{tabular}{|c|c|c|c|}
\hline
network & $Q_I$ & $Q_V$ & $Q_{I0}$ \\
\hline
\hline
~~ ACH~~ & ~~$0.15$~~ & ~~$0.10$~~ & ~~$0.16$~~ \\
\hline
~~ ACL~~ & ~~$0.15$~~ & ~~$0.12$~~ & ~~$0.16$~~ \\
\hline
~~ ACV~~ & ~~$0.11$~~ & ~~$0.16$~~ & ~~$0.12$~~ \\
\hline
~~ AHL~~ & ~~$0.33$~~ & ~~$0.04$~~ & ~~$0.37$~~ \\
\hline
~~ AHV~~ & ~~$0.16$~~ & ~~$0.12$~~ & ~~$0.17$~~ \\
\hline
~~ ALV~~ & ~~$0.16$~~ & ~~$0.11$~~ & ~~$0.18$~~ \\
\hline
~~ CHL~~ & ~~$0.31$~~ & ~~$0.13$~~ & ~~$0.32$~~ \\
\hline
~~ CHV~~ & ~~$0.09$~~ & ~~$0.09$~~ & ~~$0.12$~~ \\
\hline
~~ CLV~~ & ~~$0.12$~~ & ~~$0.11$~~ & ~~$0.12$~~ \\
\hline
~~ HLV~~ & ~~$0.32$~~ & ~~$0.07$~~ & ~~$0.33$~~ \\
\hline
\hline
~~ ACHL~~ & ~~$0.38$~~ & ~~$0.15$~~ & ~~$0.38$~~ \\
\hline
~~ ACHV~~ & ~~$0.20$~~ & ~~$0.16$~~ & ~~$0.20$~~ \\
\hline
~~ ACLV~~ & ~~$0.20$~~ & ~~$0.17$~~ & ~~$0.21$~~ \\
\hline
~~ AHLV~~ & ~~$0.39$~~ & ~~$0.14$~~ & ~~$0.39$~~ \\
\hline
~~ CHLV~~ & ~~$0.34$~~ & ~~$0.16$~~ & ~~$0.35$~~ \\
\hline
\hline
~~ ACHLV~~ & ~~$0.41$~~ & ~~$0.20$~~ & ~~$0.41$~~ \\
\hline
\end{tabular}
\caption{Normalized SNRs $Q_I$, $Q_V$ and $Q_{I0}$ for 
  network of detectors. }
\label{nett}
\end{table}

>From Table \ref{nett} and Figure \ref{f25}, several 
diagnostic features are summarized: 

\begin{description}
\item[(i)]  To realize a good sensitivity to the $I$ mode,  the  HL pair  
has a crucial role. This is due to their small separation.  Without the pair, 
we have at most $\mathrm{SNR}_I=0.20$ and $\mathrm{SNR}_{I0}=0.21$. 
Including the two detectors, the value $\mathrm{SNR}_I$ becomes more than 0.3.

\item[(ii)]  The combination AHL does not have a good sensitivity to the
$V$ mode and we obtain $Q_V=0.04$.  This is because the orientation of 
AIGO detector is specialized to achieve the best sensitivity to $I$ 
mode in combination with LIGO detectors. In fact, they are aligned to 
have large overlaps (in relation to (i)). Nevertheless, the sensitivity 
to the $V$ mode can be improved by adding the LCGT or Virgo detector. 
In contrast,  the value $Q_{I0}$ increases only $10\%$ even if we 
increase the network from AHL to ACHLV. 

\item[(iii)]  Comparing $\mathrm{SNR}_I$ with $\mathrm{SNR}_{I0}$, 
we deduce a tiny amount of statistical loss for the sensitivity to the $I$ 
mode,  caused by adding a new target parameter, $V$.  
 It is not preferable to get a significant statistical loss by dealing with
the circular polarization mode whose fraction is naively expected to be small. 
But Table \ref{nett}  indicates that even when the 
$V$ mode is added as observational targets, detection efficiency for $I$ 
mode remains almost unchanged. For three-detector networks, the 
relative loss is 
largest for CHV and reach about $24$\%. Increasing the number of detectors, 
the maximum loss is reduced to $4$\% for four-detector network, and  
further reduced to $1$\% for five-detector network. 
\end{description}

\begin{figure}[t]
\begin{center}
\epsfxsize=8cm
\epsffile{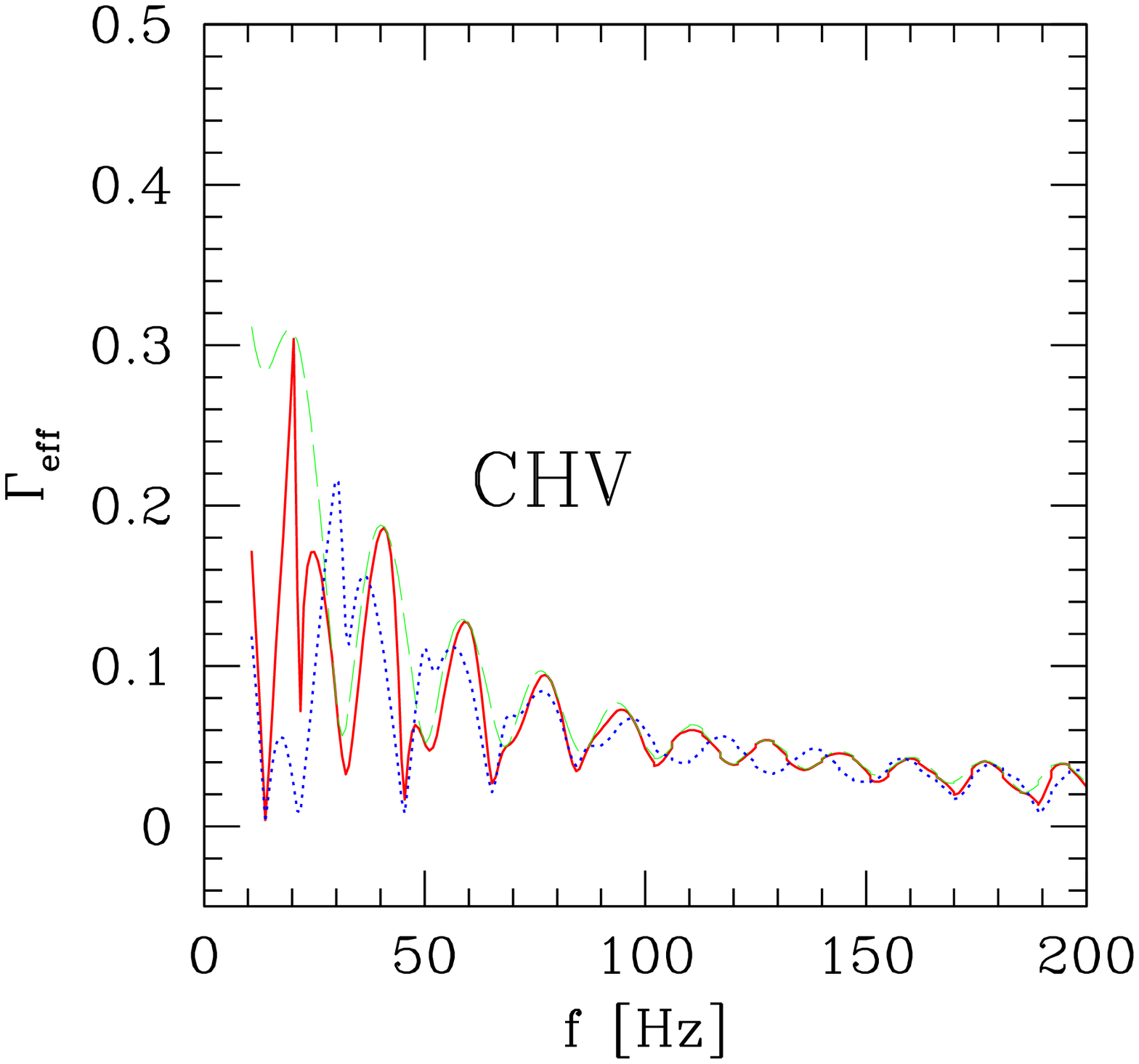}
\epsfxsize=8cm
\epsffile{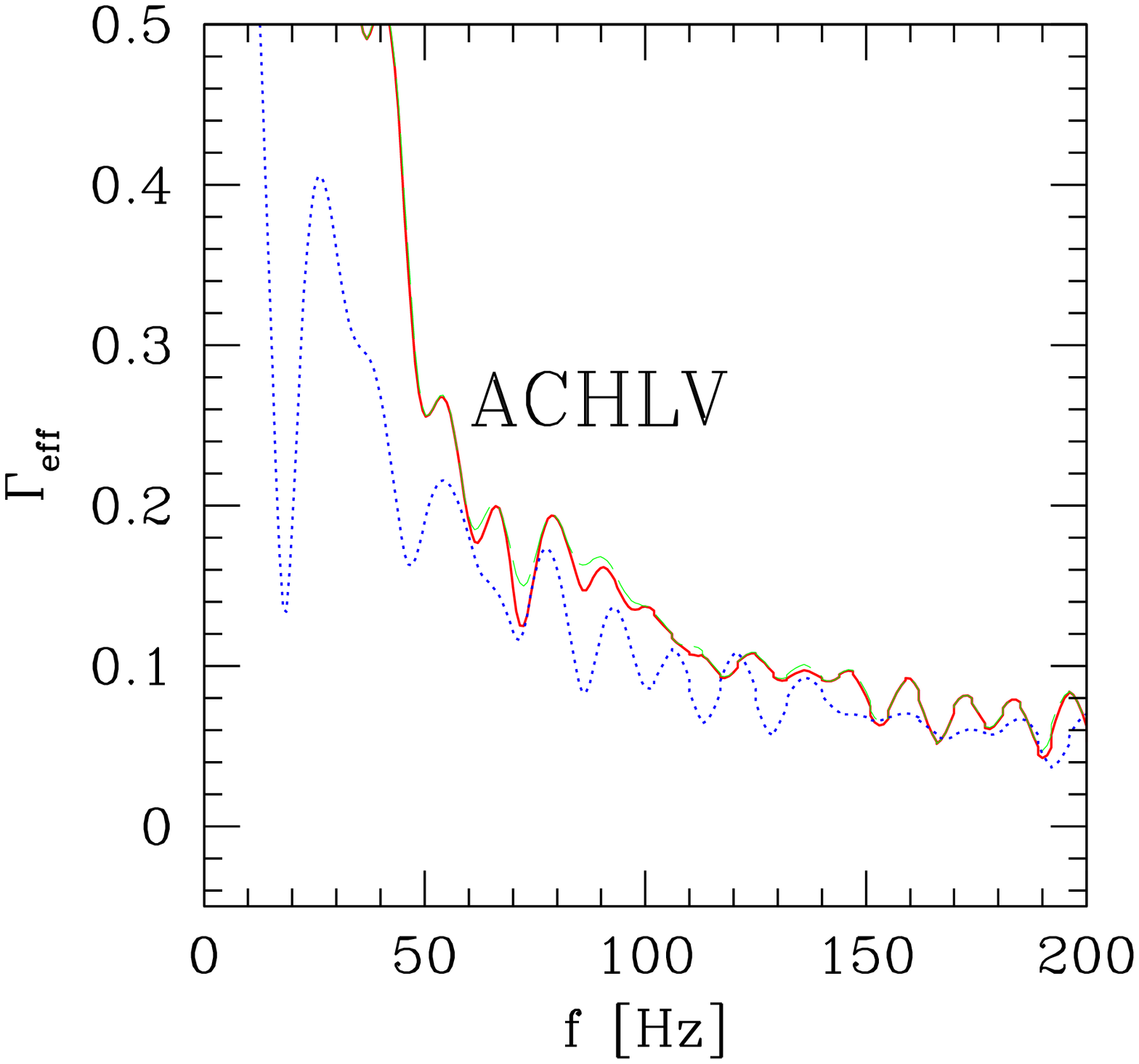}
 \end{center}

\vspace*{-0.2cm}

\caption{Left: effective overlap functions, $\Gamma_{{\rm eff},I}^2$ 
  (solid), $\Gamma_{{\rm eff},I}^2$(dotted), and 
  $\Gamma_{{\rm eff},I0}^2$ (long-dashed). Right: 
same as in left panel, but for the network of five detectors. 
Two curves for $\Gamma_{{\rm eff},I}^2$ and  $\Gamma_{{\rm eff},I0}^2$ 
are nearly overlapped. } 
\label{fig:g3_g5}

\vspace*{0.2cm}


\begin{center}
\epsfxsize=9.cm
\epsffile{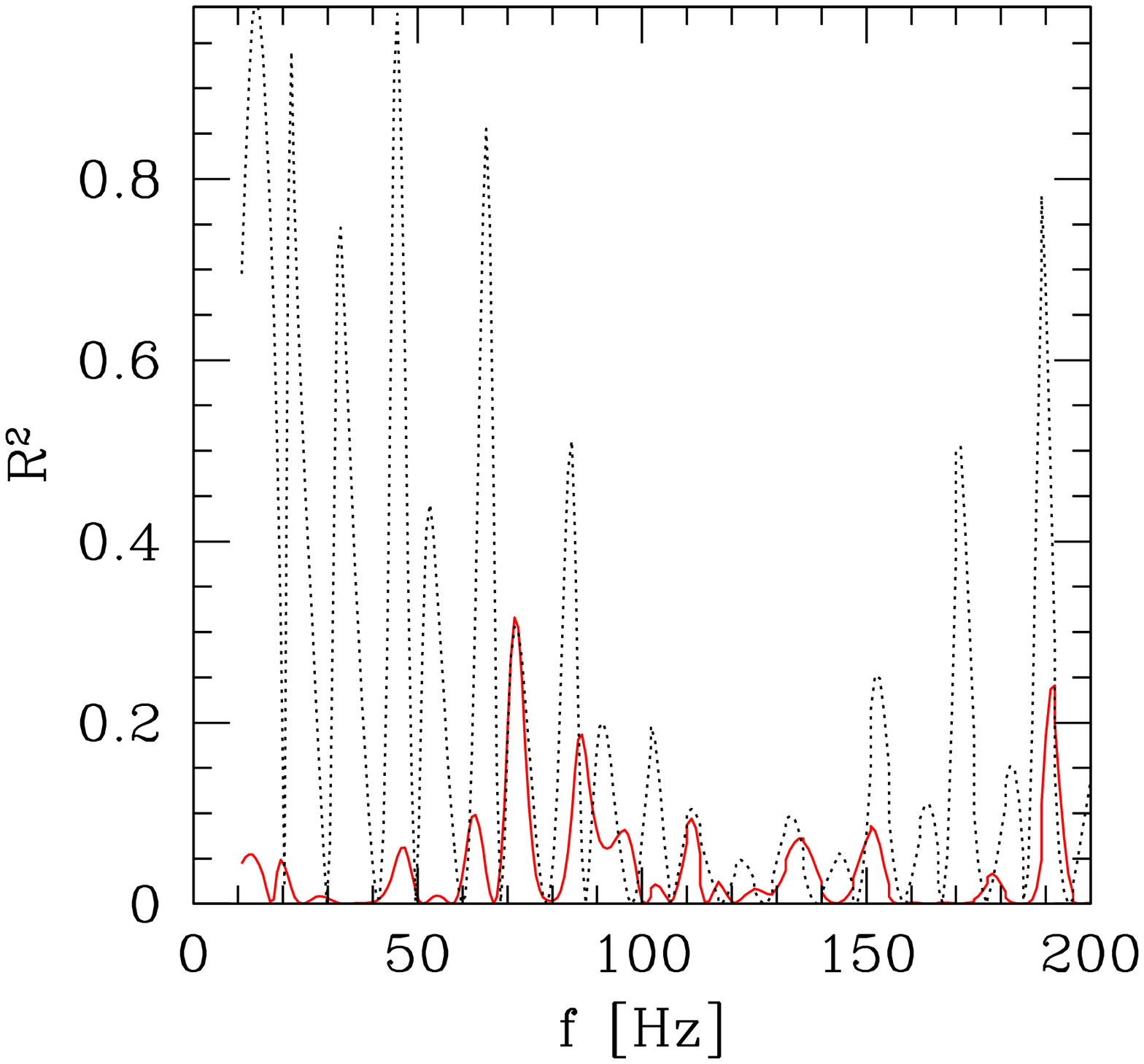}
 \end{center}

\vspace*{-0.2cm}

\caption{The function $R^2$ for the CHV (dotted curve) and the ACHLV
 networks (solid  curve).}  
\label{rr}
\end{figure}
To further give a deep insight into the diagnosis (iii), in 
Figure \ref{fig:g3_g5}, we plot 
the effective overlap functions $\Gamma_{{\rm eff},I}$, 
$\Gamma_{{\rm eff},I0}$ and $\Gamma_{{\rm eff},V}$ for specific networks 
of CHV (left) and ACHLV (right). For network of CHV, there exists 
characteristic pattern at $f>40$Hz with period $\Delta f=18$Hz. The 
main reason of this comes from the fact that all the pairs, CH, CV and HV, 
have the separation angle $\beta\sim 80^\circ$, leading to the frequency 
$\Delta f=c/(4 R_{\rm E}\sin(80^\circ/2))\sim 18$Hz. 
Focusing on the differences between $\Gamma_{{\rm eff},I}$ and
$\Gamma_{{\rm eff},I0}$, we find that while the differences are manifest at 
low-frequency in CHV system, the functions $\Gamma_{{\rm eff},I}$ 
and $\Gamma_{{\rm eff},I0}$ in ACHLV system become almost identical 
even at low-frequency. This is clearly quantified 
if we plot the ratio $R$ defined in equation (\ref{eq:def_R}). 
Figure \ref{rr} reveals that the magnitude of the ratio $R$ is 
significantly reduced for the ACHLV system, suggesting the fact that 
the statistical loss is negligibly small.  
In this respect, negligible statistical loss may be 
another merit for a network with a large number of detectors.

Finally, from numerical results given above and with a help of equation 
(\ref{norm}),  we summarize the signal-to-noise ratios, 
$\mathrm{SNR}_I$ and $\mathrm{SNR}_Q$  (not normalized ones) 
with noise spectrum of advanced LIGO.  
For the five-detector network, assuming the 
flat spectra $\Omega_{\rm GW}=\mathrm{const}$ and 
$\Omega_{\rm GW}\Pi =\mathrm{const}$, we have
\beq
\mathrm{SNR}_I=1.64 \lmk\frac{\Omega_{\rm GW}h_{70}^2}{10^{-9}}  \rmk
\lmk\frac{T_{\rm obs}}{3 \rm yr}  \rmk^{-1/2},
\quad
\mathrm{SNR}_V=0.749 \lmk\frac{\Omega_{\rm GW}h_{70}^2\Pi }{10^{-9}}  \rmk
\lmk\frac{T_{\rm obs}}{3 \rm yr}  \rmk^{-1/2}.
\eeq

\section{Summary}
\label{sec:summary}

In this paper we present prospects for measuring the Stokes
$V$ parameter of  stochastic gravitational wave backgrounds via 
the correlation analysis.  This 
parameter characterizes the asymmetry of amplitudes of right- and
left-handed waves.  As the parity transformation interchanges the two
polarization modes,   it can be regarded as the basic
observational measure to probe parity violation.  We  made detailed analyses
for the basic properties of the overlap functions $\gamma_I$ and
$\gamma_V$, especially their dependencies on geometry of detector
configurations.   

In contrast to studies only for the unpolarized $I$ mode 
(equivalently,  the energy spectrum $\Omega_{\rm GW}$),  we need to develop a
new statistical framework to deal with rich  structures 
caused by  multi-dimensionality of target parameters.   We provide an 
optimal method that will be applicable to  various problems of 
gravitational-wave backgrounds.  Based on our new method, we 
estimated sensitivities of the planned and proposed
next-generation interferometers to the  $V$ modes.   We  found that it is 
important to have a large number of detectors in order to reduce 
possible effects due to correlation between target parameters.

\bigskip
We would like to thank M. Ando, N. Kanda and M. Ohashi for useful
conversations. This work was in part supported by a 
Grant-in-Aid for Scientific Research from the Japan Society for the 
Promotion of Science (No.~18740132). 

\appendix
\section{Tensorial expansion}
\label{sec:tensor_analysis}

In this appendix we present tensorial decompositions of the overlap
functions  $\gamma_{I,V}$ (see also Ref.\cite{Flanagan:1993ix}) 
defined for pair of detectors $a$ and $b$ at positions $\vex_a$ and 
$\vex_b$.  They are expressed as
\beq
\gamma_{I,ab}(f)=\frac{5}{8\pi}\int_{S^2} d\ven  \lkk  \lnk
F_a^+F_{b}^{+*}+
F_a^\times F_{b}^{\times*} \rnk e^{iy \ven\vem} \rkk, \label{gi1}
\eeq
 and
\beq
\gamma_{V,ab}(f)=\frac{5}{8\pi}\int_{S^2} d\ven
\lkk i \lnk
F_a^+F_{b}^{\times*}-
F_a^\times F_{b}^{+*} \rnk e^{iy \ven\vem}  \rkk ,
\eeq
with $\vex_a-\vex_b=D \vem$ ($D$: distance, $\vem$:unit vector) and 
$y\equiv2\pi fD/c$.
The beam-pattern functions $F_a^P$ are written by the the polarization
tensor $\ve^P$ and the detector tensor $\ved^a$ as
\beq
F_a^P=\ved_a:\ve^P(\ven)=d_{ij}^a e^P_{ij}.
\eeq
Here the detector tensor $\ved_a$ is given by two orthonormal vectors $\veu_a$ and $\vev_a$ as $\ved_a=(\veu_a\otimes \veu_a-\vev_a\otimes \vev_a)/2$.
Therefore, the  overlap functions are formally written as
\beq
\gamma_{I}(f)=\Gamma_{I,ijkl}(f) d_{ij}^a
d_{kl}^b,~~~\gamma_{V}(f)=\Gamma_{V,ijkl}(f) d_{ij}^a d_{kl}^b. \label{tbeq}
\eeq
In these expressions we defined
\beq
\Gamma_{I,ijkl}(f)=\frac{5}{8\pi}\int_{S^2} d\ven  \lkk  
e^+_{ij }(\ven)e ^{+}_{kl}(\ven)+
e^\times _{ij }(\ven)e_{kl}^{\times}(\ven)    \rkk  e^{iy \ven\vem}
\eeq
and
\beq
\Gamma_{V,ijkl}(f)=-\frac{5i }{8\pi}\int_{S^2} d\ven  \lkk  
e^+_{ij }(\ven)e ^{\times}_{kl}(\ven)-
e^\times _{ij}(\ven)e_{kl}^{+}(\ven)    \rkk  e^{iy \ven\vem}.
\eeq
There are apparent symmetries with respect to the subscripts of the tensor
$\Gamma_{I,ijkl}$. For example, it is invariant under the replacement $i
\leftrightarrow j$ or $(i,j)\leftrightarrow (k,l) $. 
Furthermore, with using the correspondences $e^+_{ij }(-\ven) = e^+_{ij
}(\ven)$ and $e^\times _{ij}(-\ven)= -e^\times _{ij}(\ven)$ for parity
transformation, the tensor $\Gamma_{I,ijkl}$ is a real function  taking a same value at  $\vem$ and
$-\vem$. 
>From these symmetries, the tensor $\Gamma_{I,ijkl}(f)$ is
given by a combination of basic tensors $m_i$ and  $\delta_{ij}$ as
follows; 
\beqa
\Gamma_{I,ijkl}(f)&=&a_{I1}\delta_{ij}\delta_{kl}+a_{I2}(\delta_{ik}\delta_{jl}+\delta_{il}\delta_{jk})+ 
a_{I3} (\delta_{ij}m_k m_l+\delta_{kl}m_i m_j)+a_{I4} m_i m_j m_k m_l\nonumber\\
& & + a_{I5}(\delta_{ik}m_j m_l+\delta_{jk}m_i m_l+\delta_{jl}m_i
m_k+\delta_{il}m_j m_k) \label{GI},
\eeqa
with the expansion coefficients $a_{Ii}$. These coefficients  are given in the following manner. We firstly fix the direction vector $\vem=(1,0,0)$ and calculate the components $\Gamma_{I,ijkl}(f)$ for $(i,j,k,l)=(x,x,x,x), (x,x,y,y),(x,y,x,y),(y,y,y,y)$ and $(y,y,z,z)$.  Then we can solve the coefficients
$a_{Ii}$ with their five independent combinations.  After some calculation, we obtain them in terms of 
spherical Bessel functions with  argument $y$ as
\beq
a_{I1}=0, ~~a_{I2}=j_0-\frac{10}7 j_2+\frac1{14}j_4,~~a_{I3}=-\frac{20}7
j_2-\frac5{14}j_4,~~a_{I4}=\frac52 j_4,~~~~a_{I5}=\frac{15}7
j_2-\frac5{14}j_4 \label{ai}.
\eeq
A detector tensor $\ved_a$ is usually traceless
($d_{ij}^a\delta_{ij}=0$), as we measure quadrupole 
deformation of space {\it e.g} with  interfering laser beams of two
arms.  Then  the first and third terms in equation (\ref{GI}) do not provide
contribution to the overlap function $\gamma_I$. 
With angular parameters $(\beta,\sigma_1,\sigma_2)$ for a given detector pair
$(a,b)$ on a sphere with radius $R$ (see figure 1), we can set  their positions as $\vex_a=R(\cos\beta/2,0,\sin\beta/2)$ and 
$\vex_b=R(\cos\beta/2,0,-\sin\beta/2)$, since only their relative positions are relevant.
In this case we have $\vem=(0,0,-1)$, and the two unit vectors 
$\veu_a$ and $\vev_a$ are written by
\beq
\veu_a=\cos\sigma_1(\sin\beta/2,0,-\cos\beta/2)+\sin\sigma_1 (0,1,0),~~
\vev_a=-\sin\sigma_1(\sin\beta/2,0,-\cos\beta/2)+\cos\sigma_1 (0,1,0),
\eeq
while we can put
\beq
\veu_b=\cos\sigma_2(\sin\beta/2,0,-\cos\beta/2)+\sin\sigma_2 (0,1,0),~~
\vev_b=-\sin\sigma_2(-\sin\beta/2,0,-\cos\beta/2)+\cos\sigma_2 (0,1,0)
\eeq
for the second detector $b$. With plugging in these expressions into equation (\ref{tbeq}) we obtain equation (\ref{gi}).

Similarly, the tensor $\Gamma_{V,ijkl}$ is invariant with replacements
such as $i
\leftrightarrow j$, but it is asymmetric for the replacement $(i,j)\leftrightarrow
(k,l) $ or $m_i \to -m_i$.  
The tensor  is real due to the parity relation as for $\Gamma_{I,ijkl}$. 
We found that 
it is expanded with the basic tensors as 
\beq
\Gamma_{ijkl}^V=a_{V1}
(\omega_{ik}\delta_{jl}+\omega_{il}\delta_{jk}+\omega_{jk}\delta_{il}+\omega_{jl}\delta_{ik})+a_{V2}
(\omega_{ik}m_{j} m_l +\omega_{il} m_j m_k +\omega_{jk} m_i
m_l+\omega_{jl} m_i m_k) \label{lgv}
\eeq
with $\omega_{ij}\equiv \epsilon_{ijk}m_k$ ($\epsilon_{ijk}$:
antisymmetric tensor).
 In this case, the expansion coefficients $a_{Vi}$ are solved as
\beq
a_{V1}=j_1-\frac14 j_3,~~~a_{V2}=\frac54 j_3 \label{av}.
\eeq
We can derive equation (\ref{gv}) as in the case for the $I$ mode analyzed above.

When two detectors $a$ and $b$ are on a same plane, the tensor
$\omega_{ij}$ cannot have component with respect to the two dimensional
projected space to the plane, and we have identically $\gamma_V=0$ with
equations (\ref{tbeq}) and (\ref{lgv}).

Note that the tensor $\Gamma_I$ is an even function of $m_i$ but the
tensor 
$\Gamma_V$ is an odd function reflecting its handedness.
The function $\gamma_I$ is given by spherical Bessel
functions 
$j_i$ with even $i$, while the function $\gamma_V$ is with odd
$i$. 
The asymptotic behaviors of the spherical Bessel functions  are
\beq
j_n(y)\sim \frac1{y}\cos\lmk y-\frac{(n+1)\pi}2  \rmk 
\eeq
at $y\to \infty$.
In the same manner the peaks of the function $|\gamma_I|$ are at
$y\sim (N+1/2)\pi$ ($N$: natural number) and those for  $|\gamma_V|$ are at
$y\sim  N \pi$.
Therefore the zero points for $\gamma_V$ and $\gamma_I$ are offset
by $\Delta y=\pi/2$ at large $y$.

 At the low frequency limit  $y\to 0$ we have
\beq
j_n(y)\sim \frac{y^n}{(2n+1)!!},
\eeq
and the asymptotic behaviors of the overlap functions  are  $\gamma_V\to 0$ and
$\gamma_I\to   d^a_{ij} d^b_{ij}/2$.
For two traceless tensors $d^a_{ij}$ and   $d^b_{kl}$, the combination
$d^a_{ij} d^b_{ij}$ is the unique scalar quantity 
written by their tensor product.
In terms of the angular parameters in the main text, this limit is given
by $\cos^4(\beta/2)\cos(4\Delta)$.

\section{Probability distribution functions for correlation analysis}
\label{sec:PDF_for_corr}

In this appendix,  we study probability distribution functions
associated with correlation analysis, following Ref.\cite{Seto:2005qy}.    With  Fourier
space 
representation, each data
stream $s_a(f)$ is made by gravitational wave signal $H_a(f)$ and 
noise $n_a(f)$ as
\beq
s_a(f)=H_a(f)+n_a(f),
\eeq
and we define its noise spectrum 
\beq
\lla n_a^*(f) n_a(f')\rra=\frac12\delta_{\rm D}(f-f')N_a(f).
\eeq
 We assume that the correlation between noises 
are negligible (namely $\lla n_a^* n_b\rra=0$ for $a\ne b$), and the amplitude of the signal $\lla H_a(f)^* H_b(f) \rra$
is much smaller than that of  the noise   $\lla n_a(f)^* n_a(f)
\rra$. These
are  the conditions where  correlation analysis becomes very  powerful.
We divide the positive  Fourier space into frequency segments $F_v$
($i=1,...,N$) 
with their center frequencies $\{f_v\}$ and widths $\{\delta f_v\}$. 
 In each segment the width $\delta f_v$ is much smaller
than $f_v$, and the relevant quantities (e.g. $\Omega_{\rm GW}(f)$,
$\gamma_{ab}(f)$) are almost constant. But the width is
much larger than the frequency resolution $\Delta f\equiv T_{\rm obs}^{-1}$ 
($T_{\rm obs}$: observation period) so that each  segment contains 
Fourier modes as many as $\delta f_v/\Delta f\gg 1$. 

For 
correlation analysis we compress the
observational data $s_I(f)$ by summing up the products $s^*_a(f)s_b(f)$ 
 ($a\ne b$) in each segment $F_v$ as
\beq
\mu_v\equiv\sum_{f\in F_v} s^*_a(f)s_b(f),
\label{mean}
\eeq
where we omitted the apparent subscript $\{ab\}$ for the compressed
data  $\{\mu_v\}$ for
notational 
simplicity. As the noises are assumed to be uncorrelated, the statistical
mean 
$\lla \mu_v \rra$ is caused by  gravitational wave
signal. After some calculations,  we have a real value
\beq
\lla \mu_v\rra= \sum_{f\in F_v}\lla  H_a(f)^*H_b(f)\rra\simeq
\frac{8\pi}{5}   (I\gamma_{ab,I}+V\gamma_{ab,V})\frac{\delta
f_v}{\Delta f}. 
\label{mean_signal}
\eeq 
The fluctuations around the mean $\lla \mu_v\rra$ are dominated by the
noise under our weak signal approximation,  and its variance
$\sigma_v^2$ for  the real part of $\mu_v$  becomes 
\beq
\sigma_v^2=N_a(f_v) N_b(f_v) \frac{\delta f_v}{8\Delta f}.
\label{noise_variance}
\eeq
As the number of Fourier modes $\delta f_v/\Delta f$ in each segment
is much larger 
than unity, the probability distribution function (PDF) for the real 
part of  the measured value $\mu_v$ is  close to 
Gaussian distribution due to the central-limit theorem as
\beq
p({\rm Re} [\mu_v])\simeq\frac1{\sqrt{2\pi \sigma_v^2}}\exp \lkk-\frac{({\rm Re} [\mu_v]-\lla \mu_v\rra)^2}{2\sigma_v^2} \rkk. \label{pdf}
\eeq
Here we neglected the prior
information of the spectrum $\Omega_{\rm GW}(f)$ and $\Pi(f)$.
 From equations (\ref{mean_signal}) and (\ref{noise_variance}), the signal to noise ratio of each segment becomes
\beq
\mathrm{SNR}^2_v=\frac{\lla \mu_v\rra^2 }{\sigma_v^2}=\lmk
\frac{16\pi}{5}\rmk^2  {T_{\rm obs}} \lkk 2\delta f_v
\frac{ (I\gamma_{ab,I}+V\gamma_{ab,V})^2}{N_a(f)N_b(f)}   \rkk. 
\eeq
Summing up the all the segments quadratically, we get the total signal 
to noise ratio 
\beq
\mathrm{SNR}^2= \lmk \frac{16\pi}{5}\rmk^2 T_{\rm obs} \lkk2 \int_0^\infty df
\frac{ (I\gamma_{ab,I}+V\gamma_{ab,V})^2}{ N_a(f)N_b(f)}   \rkk. 
\label{single}
\eeq
This expression does not depend on the details of the
segmentation $\{F_v\}$. The same  results  can be derived by  introducing 
the optimal filter for the product $N_a^*(f) N_b(f)$ to get the highest 
signal to  noise ratio (see {\it e.g.,} Ref.\cite{Flanagan:1993ix}).

\section{Derivation of optimal SNRs for multiple detectors}
\label{sec:derivation_optimal_SNR}

The derivation outlined in Sec.\ref{subsec:multiple_correlation} is intuitive, but it is algebraically
complicated  to derive the final expressions. For example, we need to
deal with large-dimensional noise matrices ${\cal M}_{Iij}$ and ${\cal
M}_{Vij}$ that have off-diagonal components.  In this Appendix, we make a simple
explanation for the structure of equations (\ref{31}) and (\ref{32}), and derive  
useful expressions valid for arbitrary number of detectors $n$. Based on
Appendix \ref{sec:PDF_for_corr}, we consider the summed correlation signals 
$\{\mu_{vi}\}$ in a 
fixed small band $F_v$ with its bandwidth $\delta f_i$ as in equation (B3).  
Later, we will sum up all the bands to get the total SNRs.

In  actual observation, we cannot exactly  measure the expectation values 
\beq
\lla \mu_{vi}\rra=C_i b_v=\frac{8\pi}5\{\gamma_{Ii} I_+\gamma_{Vi} V\}b_v.
\eeq
Rather, the measured values $\{\mu_{vi}'\}$ fluctuate around the true
values $C_i b_v$ with variances ${\cal N}_i^2 b_v$.  Here the ratio
$b_v=\frac{\delta f_v}{\Delta f}$ is the number of frequency bin in the
band $f_v$ with the frequency 
resolution $\Delta f=T_{\rm obs}^{-1}$.  The multi-dimensional
probability distribution function $P(\{\mu_{vi}'\})$ has a form
\beq
P(\{\mu_{vi}'\})\propto
\exp\lkk -K\rkk \label{prob}
\eeq
with the kernel
\beq
K\propto \sum_{n_{t}}\frac{(\mu_{vi}'-\lla\mu_{vi}\rra)^2}{{\cal N}_i^2} b_v.
\eeq

The structure of the distribution function $P(I_e,V_e)$ for the
estimated values 
$\{I_e,V_e\}$ is obtained by the replacement
\beq
\mu'_{vi}\to \frac{8\pi}5(\gamma_{Ii} I_e+\gamma_{Vi} V_e)b_v.
\eeq
Since the expression in the large parenthesis $[\cdots]$
 in equation (\ref{prob}) becomes a quadratic function of the target parameters
$\{I_e,V_e\}$, their 
expectation values are $\{I,V\}$, and  their covariance noises matrix
is proportional to
\beq
\left( \begin{array}{@{\,}cc@{\,}}
           a_{II} & a_{IV}  \\
           a_{IV} & a_{VV}  \\ 
           \end{array} \right)^{-1}=\frac1{a_{II}a_{VV}-a_{IV}^2} \left(\begin{array}{@{\,}cc@{\,}}
           a_{VV} & -a_{IV}  \\
           -a_{IV} & a_{II}  \\ 
           \end{array} \right)
\eeq
with the matrix elements, $a_{II}=\sum_i \gamma_{Ii}^2/{\cal N}_i$, 
$a_{VV}=\sum_i \gamma_{Vi}^2/{\cal N}_i$ and 
$a_{IV}=\sum_i (\gamma_{Ii}\gamma_{Vi})/{\cal N}_i$.
Note that the off-diagonal element $a_{IV}$ has information for the
statistical correlation between $I$- and $V$-modes. We define the ratio
$R$ by
\beq
R=\frac{a_{IV}}{\sqrt{a_{II}a_{VV}}}.
\eeq
The ratio of the expectation values squared $\{I^2,V^2\}$ to the
variances of their noises are proportional to 
\beq
I^2\frac{a_{II}a_{VV}-a_{IV}^2 }{  a_{VV}}, 
\quad
V^2\frac{a_{II}a_{VV}-a_{IV}^2  }{a_{II}}.
\eeq
These expressions exactly coincide with equations (\ref{31}) and 
(\ref{32}) in the case of $n_t=3$. 
Indeed, using $Mathematica$, we confirmed  that the above 
expressions faithfully reproduced  the same results as 
obtained from the direct estimation with equation 
(\ref{eq:SNR_direct_method}), up to $n_t=8$. 
By summing up all the frequency bands and properly dealing with the
number of bins $b_v=\delta f_v/\Delta f$ (as in Appendix \ref{sec:PDF_for_corr}), we can easily evaluate the broadband SNRs for large number of detectors with
the following the expressions:
\beqa
\mathrm{SNR}_I^2&=&\lmk\frac{16\pi}{5}  \rmk^2 T_{\rm obs} \lkk
2\int_0^\infty df I^2\frac{a_{II}a_{VV}-a_{IV}^2 }{  a_{VV}} \rkk, \\
\mathrm{SNR}_V^2&=&\lmk\frac{16\pi}{5}  \rmk^2 T_{\rm obs}  \lkk
2\int_0^\infty df    V^2\frac{a_{II}a_{VV}-a_{IV}^2 }{  a_{II}} \rkk .
\eeqa
For  analysis  only
with the $I$-mode (or $\Omega_{\rm GW}$), the broadband SNR for the $I$ mode is
evaluated by 
\beq
\mathrm{SNR}_{I0}^2=\lmk\frac{16\pi}{5}  \rmk^2 T_{\rm obs} \lkk
2\int_0^\infty df {I^2 a_{II}} \rkk .
\eeq
We use this expression as a reference to analyze effects caused by
estimation of multiple parameters.
We can express $\mathrm{SNR}_I$ as
\beq
\mathrm{SNR}_I^2=\lmk\frac{16\pi}{5}  \rmk^2 T_{\rm obs} \lkk
2\int_0^\infty df I^2 a_{II} (1-R^2) \rkk .
\eeq
Therefore the ratio $R$ characterizes the loss of SNRs due to the
increase of the number of observable parameters. A similar argument holds
for the $V$ mode.

\section{Detectors on the Moon}
\label{sec:moon}

It has been discussed that, on the surface of the Moon, the
high-vacuum level and rich three-dimensional surface structure
 are suitable to build a gravitational wave interferometer with very long
armlength (see {\it e.g.,} Ref.\cite{moon}). 
Meanwhile,  the north and south poles of the Moon seem to be 
preferable and are thought to be special places for human 
activities as well as astronomical observation (see {\it e.g.,} 
Ref.\cite{ice}). 
The rotation axis of the Moon is nearly perpendicular to the ecliptic plane. 
Therefore, at the rims of craters
near the two poles, the sun-light is available during most of one Moon's
day ($\sim$30 Earth days). In contrast, the bottom of craters around the poles
is at permanent night with exceptionally stable temperature environment
around 40K, and we might obtain trapped water ice, from which we can produce 
hydrogen and oxygen (fuel for rocket engine) with electrical decomposition. 
Away from the pole areas, the surface of the Moon has severe physical 
conditions with a large temperature variation typically from $\sim 100$K 
(night) to $\sim 400$K (daytime).  Therefore, when we build detectors on the 
Moon, location near the poles would be the most natural choice. 


\end{document}